\documentclass[a4paper,11pt]{article}
\pdfoutput=1
\usepackage{jheppub}
\usepackage[T1]{fontenc}
\usepackage[utf8]{inputenc}
\usepackage{mathtools}
\usepackage{xspace}
\usepackage{xcolor}
\usepackage{booktabs}
\usepackage{multirow}

\newcommand{\qb}{\bar{q}}
\newcommand{\qp}{q'}

\newcommand\id[1]{#1^\text{id.}}

\newcommand{\XTZ}[2]{\mathcal{#1}_3^{0,\text{id.}#2}}

\newcommand{\MF}{{\rm MF}}
\newcommand{\as}{\alpha_s}
\newcommand{\e}{\epsilon}
\newcommand{\B}{\rm B}
\newcommand{\rto}{\leftarrow}

\newcommand{\MFK}[2]{\Gamma^{(#1)}_{#2}}

\newcommand{\der}{\mathrm{d}}
\newcommand{\GeV}{\mathrm{GeV}}
\newcommand\ptj{p_{T,j_1}}
\def\hsig{\hat{\sigma}}
\def\LO{{\rm LO}}
\def\NLO{{\rm NLO}}
\def\NNLO{{\rm NNLO}}
\def\R{{\rm R}}

\def\V{{\rm V}}
\def\S{{\rm S}}

\def\T{{\rm T}}

\def\q{\mathcal{Q}}

\def\JET{J}

\preprint{{\raggedleft%
ZU-TH 25/24 
CERN-TH-2024-062
}}

\title{QCD predictions for vector boson plus hadron production at the LHC}

\author[a]{S.~Caletti,}
\author[a,b]{A.~Gehrmann--De Ridder,}
\author[c]{A.~Huss,}
\author[a]{A.~Rodriguez Garcia,}
\author[d]{G.~Stagnitto}

\affiliation[a]{Institute for Theoretical Physics, ETH, CH-8093 Z\"urich, Switzerland}
\affiliation[b]{Department of Physics, University of Z\"urich, CH-8057 Z\"urich, Switzerland}
\affiliation[c]{Theoretical Physics Department, CERN, CH-1211 Geneva 23, Switzerland}
\affiliation[d]{Universit\`{a} degli Studi di Milano-Bicocca \& INFN, Piazza della Scienza 3, I-20126 Milano, Italy}

\emailAdd{scaletti@phys.ethz.ch}
\emailAdd{gehra@phys.ethz.ch}
\emailAdd{alexander.huss@cern.ch}
\emailAdd{adrianro@phys.ethz.ch}
\emailAdd{giovanni.stagnitto@unimib.it}

\abstract{
The identification of a hadron in the final state of hadron-collider events that feature a leptonically decaying vector boson can provide essential information on the parton content of the colliding protons.
Moreover, the study of hadrons inside jets can provide deeper insights into the fragmentation dynamics.
We provide theoretical predictions for specific observables involving either the production of a $Z$ boson in association with light charged hadrons inside a jet or the production of a $W$ boson together with a charmed hadron.
We present results for various fragmentation functions and compare our predictions with measurements by LHCb and ATLAS at $\sqrt{s}=13$\ TeV.
Our predictions are obtained using the antenna subtraction formalism which has been extended to cope with infrared singularities associated to the fragmentation processes in a hadron-collider environment at NLO accuracy. 
}

\begin{document} 
\maketitle


\section{Introduction}
\label{sec:intro}

With the rapidly growing data set from the LHC and the reduction of both statistical and systematic uncertainties, more and more observables are becoming accessible for precision studies.
One such frontier comprises processes with identified hadrons from light or heavy quark flavours.

Identified hadron production is described in perturbative QCD through the production of partons 
(quarks and gluons) which subsequently fragment into hadrons. This parton-to-hadron transition is a non-perturbative process which can be parametrized in terms of process-independent fragmentation functions (FFs). The latter describe the probability of a parton fragmenting into a hadron carrying some fraction of its momentum.
These FFs fulfil Altarelli-Parisi
evolution equations in their resolution scale~\cite{Altarelli:1977zs}, which are
in complete analogy to the evolution of parton distributions functions (PDFs) in the nucleon.
%

Identified hadron production can arise in two categories of collider observables: 
{\em inclusive} observables, which are differential in the hadron momentum 
but inclusive in the kinematics of all other final state particles;
{\em exclusive} observables, where the hadrons are identified in events that are 
characterized by requiring the additional presence of jets and/or gauge bosons in the final state.

One-particle inclusive cross sections have been measured for a 
variety of hadron species in electron-positron, lepton-hadron and hadron-hadron collisions. 
On the theory side, using the factorization properties of QCD, the identified hadron production cross section can be written as the convolution of a process-dependent coefficient function for the production of partons with a universal fragmentation function for the parton-to-hadron transition.
The coefficient functions, corresponding to the parton-level cross section can be 
calculated in perturbative QCD to the desired orders. 
At present, these coefficient functions are known to NLO for
hadron-hadron collisions~\cite{Aversa:1988vb} and to NNLO
 for $\mathrm{e}^+\mathrm{e}^-$ annihilation~\cite{Rijken:1996ns,Mitov:2006ic} and lepton-proton
scattering~\cite{Bonino:2024qbh}. 


The recent focus on exclusive observables is largely driven by precision studies of vector boson production in association with identified hadrons at the LHC, with the presence of leptons from the vector boson decay providing a clean experimental signature.
These exclusive observables can then provide essential information on the quark flavour decomposition of the colliding protons: for instance, $W$-boson plus charmed hadron processes are important handles on the strange content of the proton~\cite{ATLAS:2014jkm,CMS:2018dxg,CMS:2021oxn,CMS:2023aim,ATLAS:2023ibp}.

%

Other exclusive observables require the presence of a hadron inside a jet~\cite{Arleo:2013tya}.
These processes can provide significant constraints on fragmentation functions.
Recently, observables involving the production of a $Z$-boson in association with light hadrons inside a jet have been performed by the LHCb collaboration~\cite{LHCb:2022rky}.

Theory predictions for any of these exclusive observables require the use of a
parton-level event generator implementing all parton-level contributions to a given order and
applying the experimental kinematical requirements on the final state event selection.
%
%
For the computation of observables related to processes with hadrons and involving parton-to-hadron fragmentation functions, any subtraction method originally developed for jet production needs to be extended: the subtraction procedure must keep track of the identified parton momentum fraction in the unresolved
emissions, which is usually integrated over for purely jet observables.
Such extensions are available at
NLO for dipole subtraction~\cite{Catani:1996vz} and FKS subtraction~\cite{Frederix:2018nkq}.
At NNLO, recent work towards the description of 
fragmentation processes yielded results for heavy hadron production in top quark
decays~\cite{Czakon:2021ohs} in the residue subtraction
scheme as well as for photon fragmentation~\cite{Gehrmann:2022cih,Chen:2022gpk} 
in the antenna subtraction scheme. An extension of the antenna subtraction formalism to incorporate 
arbitrary hadron fragmentation processes at lepton colliders has been outlined in~\cite{Gehrmann:2022pzd}.
Very recently, an interface to MG5\_aMC@NLO~\cite{Alwall:2014hca} to
compute exclusive processes with fragmentation at NLO through the usage of a
hybrid subtraction scheme has been presented~\cite{Liu:2023fsq}.
%
Other theoretical developments concern the interface of fixed-order NLO calculations to parton showers for processes with identified hadrons in the final state:
for instance, NLO+PS predictions for $W$ plus charmed hadron production have been produced in~\cite{Bevilacqua:2021ovq,FerrarioRavasio:2023kjq}.
 

In this paper, we consider theory predictions for $Z$-boson plus light hadron and $W$-boson plus heavy hadron production.
We present the essential ingredients of the antenna formalism extended to cope with infrared singularities associated to fragmentation processes at NLO accuracy.
This work is a first step towards obtaining NNLO predictions for vector boson plus identified hadron production at the LHC.


The paper is structured as follows.
In section~\ref{sec:theory}, we present the general framework describing hadron fragmentation processes for hadron collider observables up to NLO level.
In section~\ref{sec:subtraction_V+H} we give explicit expressions for the subtraction terms and their integrated counterparts needed for the computation of hadron production in association with a $Z$ or $W$ boson.
As a first application of our formalism, in sections~\ref{sec:Z+h} and~\ref{sec:W+D} we present theoretical predictions using specific parametrizations
for the fragmentation functions and perform in-depth comparisons with LHCb and ATLAS data at 13\ TeV.
Finally we summarize our findings in Section~\ref{sec:conclusions}. 

\section{Identified hadron production at hadron colliders}
\label{sec:theory}

We begin by recollecting the main features of the different ingredients that enter the computation of hadron-collider observables associated to the production of identified hadrons at NLO.

We consider a generic proton--proton to hadron process of the form: 
\begin{equation}
	\mathrm{p} + \mathrm{p} \to h(K_h) + X \, (\, + \mathrm{jets} \,) \,,
\end{equation}
where the hadron $h$ with momentum $K_h$ may or may not be inside a jet. The fully differential cross section for a process of this kind can be written in a factorized form
\begin{equation}\label{eq:sigmaH}
\der\sigma^h = \sum_{i,j} \sum_{p} \int  \frac{\der \xi_1 }{\xi_1}\frac{\der \xi_2}{\xi_2} \der\eta\, f_i(\xi_1,\mu_F^2)f_j(\xi_2,\mu_F^2)
D^h_{p}(\eta,\mu_D^2)\,\der\hsig^{i,j}_{p}(\eta,\mu_{F}^2 ,\mu_D^2) \,,
\end{equation}
where $f_{i,j}$ are the parton distribution functions (PDFs) with momentum fractions $\xi_{1,2}$ and $D^h_p$ denotes the fragmentation function (FF) describing the transition of a parton $p$ into a hadron $h$ carrying the momentum fraction $\eta = {K_h}/{k_p}$.
After mass factorization, both the PDFs and FF acquire a dependence on the factorisation ($\mu_F$) and fragmentation scale ($\mu_D$), respectively, which compensates against the corresponding scale dependence of the hard scattering cross section $\der\hsig^{i,j}_{p}(\eta,\mu_{F}^2 ,\mu_D^2)$.
A sum over the initial-state ($i$, $j$) and final-state ($p$)  partons is performed.
Finally, the one-parton exclusive cross section $\der\hsig^{i,j}_{p}$ is calculated perturbatively as a series expansion in $\alpha_s$ and reads
\begin{equation}
	\label{eq:sigmaP}  \der\hsig_p(\eta,\xi_{1,2})
	= \der\hsig^{\LO}_p(\eta,\xi_{1,2}) + \left( \frac{\as}{2\pi} \right)
	\der\hsig^{\NLO}_p(\eta,\xi_{1,2}) + \left( \frac{\as}{2\pi} \right)^2
	\der\hsig^{\NNLO}_p(\eta,\xi_{1,2}) + \ldots\,,
\end{equation}
where we specify the fragmenting parton denoted as $p$ but leave the initial state $(i,j)$ implicit.
In the following, we describe how the parton-level cross section given in eq.~\eqref{eq:sigmaP} is computed up to NLO.
We closely follow the notation adopted in~\cite{Gehrmann:2022pzd}, where the formalism
needed to incorporate arbitrary hadron fragmentation processes for lepton-collider observables
has been outlined up to NNLO.

For a specific initial state denoted as $(i,j)$ and a fragmenting parton denoted as $p$ in eq.~\eqref{eq:sigmaH},
the leading order (LO) cross section is defined as the integration over the $n$-particle phase space of the tree-level Born partonic cross section as
\begin{equation}
	\der\hsig_{p}^{{\rm LO}}(\eta,\xi_{1,2}) = \int_{n} \der\hsig_{p}^{\B}(\eta,\xi_{1,2})\,,
\end{equation}
with, 
\begin{equation}
\label{eq:Bxs}
	\der\hsig_{p}^{\B}(\eta,\xi_{1,2}) = {\cal N}_{\B}\,
	\der\Phi_{n}(k_{1},\ldots,k_{n};\xi_1P_1,\xi_2P_2)\,
	\frac{1}{S_{{n}}}\,M^{0}_{n}(k_{1},\ldots,k_{n})\,
	J^{(m)}_n(\{k_1,\ldots,k_n\}_n, \eta k_p)\,.
\end{equation}
Here, ${\cal N}_{\B}$ denotes the normalisation factor at Born-level, $S_{{n}}$ a symmetry
factor related to the final-state particles, $M^{0}_{n}$ the tree-level matrix element for $2\to n$ scattering, and $\der\Phi_{n}$ the phase space for a
$n$-parton final state with total four-momentum $Q=\xi_1P_1 + \xi_2P_2$.
Throughout the remainder of the paper, we will define the initial-state partons $(i,j)$ as $(1,2)$,
use $p_{1,2}=\xi_{1,2}P_{1,2}$, and leave the PDF momentum fractions and the flavour content of the initial-state partons implicit.

The jet function $J^{(m)}_n$ in eq.~\eqref{eq:Bxs} implicitly describes how the jet cross sections are computed.
In particular, it includes the jet clustering algorithm and application of cuts on jet observables in case their reconstruction is part of the process definition. Compared to standard jet functions, the jet function $J^{(m)}_n$ here requires modification for the computation presented in this work:
It depends explicitly on the momentum fraction $\eta$ relating the momentum $k_{p}$ of the identified parton and the momentum of the identified 
hadron $K_h=\eta k_p$.
This enables to define and apply cuts on observables that depend explicitly
on the momentum fraction $\eta$ carried by the hadron. 

At higher orders, infrared divergences due to the emission of soft and/or collinear radiation appear in the parton-level cross section. For sufficiently inclusive observables these divergences are guaranteed to cancel between real and virtual contributions. 
A subtraction method is needed however, in order to regulate these divergences at intermediate stages of the calculation.
In the antenna subtraction formalism, used in this work, subtraction terms are constructed with products of antenna functions and reduced matrix elements of lower multiplicity. The antenna functions capture all unresolved radiation between a pair of hard radiators, thus reproducing the behaviour of the matrix element in the singular limits.
The integrated subtraction terms are then obtained by analytically integrating the antenna functions over their respective 
factorised phase space. Those integrated subtraction terms are then added back at the virtual level in order to cancel the explicit poles in the virtual matrix elements and additional mass-factorisation counterterms present in a hadron-collider setup.
Most of the ingredients needed to deal with the computation of hadron-collider jet observables at NNLO using the antenna formalism~\cite{Currie:2013vh} can also be applied here in the context of identified hadron production.
However, the explicit identification of a final-state parton in $\der\hsig^{i,j}_{p}$ spoils the cancellation mechanism of collinear divergences between real and virtual contributions.
For instance, a hard object composed of a single quark or a collinear quark--gluon cluster are distinguished as separate objects in this case.
The associated collinear divergence remains uncancelled and is instead absorbed into the bare fragmentation function through mass factorization counterterms.
This renormalization procedure introduces a dependence on an arbitrary fragmentation scale $\mu_D$ at which this subtraction is performed.
In the context of the application of the
antenna subtraction formalism, this procedure has been presented in~\cite{Gehrmann:2022pzd}.

To account for the presence of an identified particle, and to guarantee that the hadron-level cross section can be computed as a convolution of the parton-level cross section and the parton-to-hadron fragmentation function as
given in eq.~\eqref{eq:sigmaH}, one needs to keep track of the momentum fraction of the fragmenting parton at all stages of the computation of the short-distance cross section. As a consequence, the construction of subtraction terms at unintegrated and integrated level within the antenna subtraction formalism must be modified accordingly.

In general, these subtraction terms will involve so-called fragmentation antennae, 
where an identified parton is tracked and its momentum dependence is kept explicitly. 
The integrated subtraction terms will involve the integration of these fragmentation antennae over the relevant phase spaces while retaining their dependence on the momentum fraction of the fragmenting parton. 
After integrating over all kinematical variables except the momentum fraction of the identified parton, these integrated fragmentation antennae will have the proper structure to be combined with virtual contributions and mass-factorisation 
counterterms associated with final-state collinear divergences of the identified parton.
The analytic pole cancellation occurs before the convolution with the fragmentation function has taken place and the finite remainder can be evaluated numerically using a parton-level event generator, like \textsc{NNLOjet}~\cite{Gehrmann-DeRidder:2015wbt}.


In the remainder of this section we present the general structure of the unintegrated and integrated subtraction terms needed to account for the presence of identified-hadron production in a hadron-collider environment at NLO.

\subsection{Subtraction at next-to-leading order}
At NLO level, the one-parton exclusive cross section as given in eq.~\eqref{eq:sigmaP} comprises two separately divergent contributions: the real-emission corrections containing implicit soft and collinear divergences and the virtual loop-corrections with explicit divergences regulated as $1/\varepsilon$ poles in dimensional regularisation. These two types of corrections feature different particle multiplicities, thus requiring a separate numerical integration.
The antenna subtraction scheme at NLO level is used to construct a real subtraction term $\der\hsig_{p}^{\S}$ and a virtual subtraction term $\der\hsig_{p}^{\T}$ in order to deal with these divergences in the intermediate stages of the computation.
As a consequence, for a specified identified parton $p$,
the $\eta$ dependent parton-level cross section $\der\hsig_{p}^{\NLO}(\eta)$
in eq.~\eqref{eq:sigmaP} may be written as:
\begin{equation}\label{eq:subNLO}
	\der\hsig_{p}^{\NLO}(\eta) =
	\int_{n+1} \left[ \der\hsig_{p}^{\R}(\eta) - \der\hsig_{p}^{\S}(\eta) \right]
	+ \int_{n} \left[ \der\hsig_{p}^{\V}(\eta) - \der\hsig_{p}^{\T}(\eta) \right]\,,
\end{equation}
with the two integrals over each particle multiplicity $n$ and $n+1$ being separately finite and thus suitable for a numerical implementation.

\subsubsection{Subtraction at real level}

The real partonic cross section $\der\hsig_{p}^{\R}$ has the same structure as the Born-level cross section
in eq.~\eqref{eq:Bxs} with an additional parton and a corresponding real-level normalisation factor.
It is composed of several terms that cover all the possibilities of having one of its $n+1$ partons becoming unresolved
with a further decomposition according to their colour structure.
Similarly, we decompose the real subtraction term $\der\hsig_{p}^{\S}$ with an identified parton $p$ into a sum of contributions
in which parton $j$ can become unresolved
\begin{equation}\label{eq:sigSNLO}
	\der\hsig_{p}^{\S} = \sum_j \der\hsig_{p,j}^{\S}\,.
\end{equation}
The term $\der\hsig_{p,j}^{\S}$ covers all relevant colour orderings and can be further split into two parts:
\begin{equation}
	\der\hsig_{p,j}^{\S} = \der\hsig_{p,j}^{\S,\text{non-id.}p} +  \der\hsig_{p,j}^{\S,\text{id.}p} \,.
\end{equation}
Here, the subtraction term denoted as $\der\hsig_{p,j}^{\S,\text{non-id.}p}$ deals with contributions in which the unresolved parton $j$ is not colour connected to the identified parton $p$. This type of term is constructed using the standard antenna formalism with conventional NLO antenna functions $X^{0}_{3}$.
In addition, the jet function ensures that the fragmenting particle is hard, i.e.\ cannot go unresolved.
Since their construction is well established, we will not discuss $\der\hsig_{p,j}^{\S,\text{non-id.}p}$ further here.

The subtraction term $\der\hsig_{p,j}^{\S,\text{id.}p}$ is instead composed of contributions
in which the unresolved parton $j$ is colour connected to the identified parton $p$.
In this case, the dependence on the fragmentation momentum fraction carried by the identified parton must be retained explicitly.
For the case of initial-state parton $1$ being the colour-connected partner of $p$, a generic real subtraction term of this type takes the following form:
\begin{eqnarray}
	\der\hsig_{p,j}^{\S,\text{id.}p} &=& {\cal N}_{\R}
	\der\Phi_{n+1}(k_{1},\ldots,k_p,\ldots,k_{n+1};p_1,p_2)\, \frac{1}{S_{{n+1}}}\,X_3^{0,\text{id.}p}(p_1;k_j,\id{k}_p)
	\nonumber \\ && \times\,
	M^{0}_{n}(k_{1},\ldots,\tilde{k}_p,\ldots,k_{n+1};xp_1,p_2)\,
	J(\{\ldots,\tilde{k}_p,\ldots\}_n,\eta\,z\,\tilde{k}_p)\,,
	\label{eq:SNLOxs}
\end{eqnarray}
where ${\cal N}_{\R} = {\cal N}_{\B}\,\overline{C}(\e)/C(\e)$, with
\begin{equation}\label{eq:Ceps}
	C(\e) = \frac{(4\pi e^{-\gamma_E})^\e}{8\pi^2}\,,\quad
	\overline{C}(\e) = (4\pi e^{-\gamma_E})^\e\,.
\end{equation}
In this equation $X_3^{0,\text{id.}p}$ is a standard three-parton initial--final antenna function involving the identified particle with momentum denoted as $k_p$. The momentum of the identified particle after an $n+1 \to n$ mapping, as it enters in the reduced matrix element, is denoted as $\tilde k_{p}$.
The information on the momentum fraction $z$ must be retained through the mapping such that the hadron momentum of the subtraction term is given by $k_h=\eta z \tilde k_{p}$.
With the common definition of the momentum fraction $x$ related to initial state emissions (with $p_1$ denoting the momentum of an initial state parton),
\begin{equation}
		x = \frac{Q^2}{2p_1 \cdot q}\,, \quad q^2=(p_q -k_p-p_1)^2=-Q^2\,, \
\end{equation}\\
in the case of initial--final kinematics, the momentum fraction $z$ can be chosen as:
\begin{equation}
	\label{eq:NLOmap}  
		z = x \frac{(k_p-p_1)^2}{Q^2} =\frac{s_{1p}}{s_{1p}+s_{1j}}\,, \\
\end{equation}
to satisfy the desired properties.
In particular, the momentum entering the jet function at this level is $\eta z \tilde k_p$ as given in eq.~\eqref{eq:SNLOxs}.

%
%
\subsubsection{Subtraction at virtual level}
\label{subsubsec:theorysection_virtual}
The virtual-level subtraction term for the one-particle exclusive cross section in
eq.~\eqref{eq:subNLO} is given by the combination of the integrated real-level subtraction terms and mass factorisation counterterms as:
\begin{eqnarray}
\label{eq:sigma_T}
\der\hsig_{p}^{\T}(\eta)=-\int_1 \der\hsig_{p}^{\S}(\eta) -\der\hsig_{p}^{\MF}(\eta).
\end{eqnarray}
Focussing on the first term of this equation,  
the integration of standard antenna functions with initial--final kinematics is expanded using the phase space factorisation presented in~\cite{Daleo:2006xa}. Here we focus on the integration of the fragmentation antennae containing the identified particle as given in eq.~\eqref{eq:SNLOxs}. The first step is to factorise the initial--final phase space and include the explicit integration over the momentum fraction $z$. This results in
\begin{eqnarray}
\label{eq:fac_phase_space}
	\der\Phi_{n+1}(k_{1},\ldots,k_p,k_j,\ldots,k_{n+1};p_1,p_2) &=&
	\der\Phi_{n}(k_1,\dots,\tilde{k}_p,\dots,k_{n+1};xp_1,p_2)\frac{\der x}{x} \der z \nonumber\\
	&\times&\frac{Q^2}{2\pi}\der\Phi_{2}(\id{k}_p,k_j;p_1,q) \delta \Big(z- \frac{s_{1p}}{s_{1p}+s_{1j}}\ \Big). \;
\end{eqnarray}
Similarly to the photon fragmentation case~\cite{Gehrmann:2022cih}, we define the integrated fragmentation antenna function denoted as $\XTZ{X}{p}(x,z)$ over the two-parton phase space as present in eq.~\eqref{eq:fac_phase_space} as:
\begin{eqnarray}
	\XTZ{X}{p}(x,z) &=& \frac{1}{C(\epsilon)}\int {\rm d}\Phi_{2}(\id{k}_p,k_j;p_1,q) \frac{Q^2}{2\pi}  \,X_3^0(p_1;k_j,\id{k}_p)  \delta\Big( z -  \frac{s_{1p}}{s_{1p}+s_{1j}} \Big) \nonumber  \\
	&= &\frac{Q^2}{2} \frac{e^{\gamma_E \epsilon}}{\Gamma(1-\epsilon)} \left(Q^2\right)^{-\epsilon} \mathcal{J}(x,z) \, X_{1,jp}^{0,\text{id.} p}(x,z)\,.
	\label{eq:intX30IFfrag}
\end{eqnarray}
Note that no integration is needed at this stage, i.e.\ just an expansion in distribution is required here.
The Jacobian factor present in eq.~\eqref{eq:intX30IFfrag} is given by
\begin{equation}
	\mathcal{J}(x,z) = (1-x)^{-\epsilon} x^{\epsilon} z^{-\epsilon} (1-z)^{-\epsilon} \,.
	\label{eq:JacPhi2}
\end{equation}

The latter factor arises from writing the two-body phase space integral as a single integral over $z$. The product of the $x$- and $z$-dependent Jacobian and antenna functions in eq.~\eqref{eq:intX30IFfrag} leads to terms of the form $(1-x)^{-1-\varepsilon}$ and $(1-z)^{-1-\varepsilon}$ which regulate the end-point soft divergences. These can be expanded in term of distributions according to
\begin{equation}
	(1-y)^{-1-k\varepsilon}= -\frac{1}{k\varepsilon}\delta(1-y)+\sum_n \frac{(k\varepsilon)^n}{n!} \mathcal{D}_n ( y) \,, 
\end{equation}
with plus-distributions of the form
\begin{equation}
	\mathcal{D}_n ( y) =\Big( \frac{\log^n(1-y)}{1-y} \Big)_+ \,.
\end{equation}
Using the integrated form of the fragmentation antenna as given in eq.~\eqref{eq:intX30IFfrag} the integrated subtraction term for an identified parton denoted as $p$ with momentum $k_p$ before the mapping, is given by
\begin{eqnarray}
	\int_1 \der\hsig_{p,j}^{\S,\text{id.}p} &=& {\cal N}_{\V}\,
	\int \frac{\der x}{x} \der z\,\der\Phi_{n}(k_1,\dots,\tilde{k}_p,\dots,k_{n+1};\q)\,
	\frac{1}{S_{n}} {\cal X}_3^{0,\text{id.} p}(x,z)
	\nonumber \\ && \times
	\,M^{0}_{n}(k_{1},\ldots,\tilde{k}_p,\ldots,k_{n+1};xp_1,p_2)\,
	J(\{\ldots,\tilde{k}_p,\ldots\}_n,\eta\,z\,\tilde{k}_p)\,,
	\label{eq:intsigS}
\end{eqnarray}
with ${\cal N}_{\V} = {\cal N}_{\R} C(\e) = {\cal N}_{\B}\,\overline{C}(\e)$.

To achieve explicit pole cancellation at virtual level with an $n$-particle configuration as formulated in eq.~\eqref{eq:subNLO}, we further need to include mass-factorisation counterterms to the integrated subtraction terms~\eqref{eq:intsigS}.
In the case of hadron-collider observables with an identified parton in the short distance cross section,
those mass-factorisation counterterms will be needed for regulating collinear divergences
coming from two different origins: initial-state collinear divergences associated with PDFs and final-state collinear singularities involving an identified parton that are associated with a FF.

With that in mind, one needs to consider mass-factorisation counterterms which are built with 
splitting kernels depending on both the initial-state momentum fraction $x$ as well as the final-state momentum fraction of the fragmenting parton $z$:
For the one identified parton denoted as $p$, the general structure of this mass-factorisation counterterm has the following 
form: 
\begin{eqnarray}
\label{eq:mass_fac_term}
	\der\hsig^{\MF,\text{id.}p}_{p, (if)}(\eta) &=& - {\cal N}_{\V}\, 
	\int \frac{\der x}{x}  \der z \,\der\Phi_{n}(k_1,\dots,\tilde{k}_p,\dots,k_{n};xp_1,p_2)\,
	\nonumber \\ && \times 
	{ \Gamma}^{(1)}_{rs;kp} (x,z) \,\frac{1}{S_{n}}  M^{0}_{n}(k_{1},\ldots,\tilde{k}_p,\dots,k_{n};xp_1,p_2)
	\nonumber \\ && \times  
	J(\{k_{1},\ldots,\tilde{k}_p,\ldots,k_{n}\}_n, \eta\,z\,\tilde{k}_p)\,. 
\end{eqnarray}
We define the $x$- and $z$-dependent mass-factorisation kernel $\Gamma^{(1)}_{rs;kp} (x,z)$ as the sum
of two terms depending either on $x$, and the factorisation scale $(\mu_F)$ or on $z$ 
and the fragmentation scale $(\mu_D)$  individually. It reads: 
\begin{eqnarray}
	\Gamma_{rs;kp}^{(1)}(x,z)=\delta (1-z)  \delta_{ps}  \mu_F^{-2\epsilon}  \MFK{1}{k \rto r}(x) + \delta (1-x) \delta _{kr} \mu_D^{-2\epsilon}\MFK{1}{p\rto s}(z)\,.
	\label{eq:gammageneric}
\end{eqnarray}
In this formula, the fragmenting parton is denoted by $p$ and there is in addition an implicit sum over $r$ and $s$, respectively denoting the particle types of the initial- and final-state partons.
The general form of the mass-factorisation counterterms needed 
in space-like and time-like kinematics is well known. In the context of the antenna subtraction formalism, explicit expressions have been presented in~\cite{Currie:2013vh} and \cite{Gehrmann:2022pzd} respectively.

Combining the expressions for the integrated subtraction term given in eq.~\eqref{eq:intsigS} and for the mass factorisation
counterterm as given in eq.~\eqref{eq:mass_fac_term}, one can express the virtual subtraction term given in eq.~\eqref{eq:sigma_T} in terms of fragmentation dipoles denoted as $\boldsymbol{J} _{2} ^{(1),\text{id.}p}$ below:
 
\begin{eqnarray}
\der\hsig_{p}^{\T\text{id.}p}(\eta) &=& -{\cal N}_{\V}\,
        \int \frac{\der x}{x} \der z\,\der\Phi_{n}(k_1,\dots,\tilde{k}_p,\dots,k_{n+1};\q)\,
        \frac{1}{S_{n}}
         \Bigg [{\cal X}_3^{0,\text{id.} p}(x,z) -\Gamma_{rs;kp}^{(1)}(x,z)\Bigg] \nonumber \\&&
        \times
        \,M^{0}_{n}(k_{1},\ldots,\tilde{k}_p,\ldots,k_{n+1};xp_1,p_2)\,
        J(\{\ldots,\tilde{k}_p,\ldots\}_n,\eta\,z\,\tilde{k}_p)\,,
        \nonumber\\
        &=& -{\cal N}_{\V}\,\int \frac{\der x}{x} \der z\,\der\Phi_{n}(k_1,\dots,\tilde{k}_p,\dots,k_{n+1};\q)\,
        \frac{1}{S_{n}} \times \boldsymbol{J} _{2} ^{(1),\text{id.}p}(p_1,\tilde{k}_p,x,z) \nonumber \\&&
        \times
        \,M^{0}_{n}(k_{1},\ldots,\tilde{k}_p,\ldots,k_{n+1};xp_1,p_2)\,
        J(\{\ldots,\tilde{k}_p,\ldots\}_n,\eta\,z\,\tilde{k}_p)\,. \nonumber
        \\
        \label{eq:J21def}
\end{eqnarray}

\section{Infrared structure of vector boson and hadron production}
\label{sec:subtraction_V+H}
In this section, we provide explicit expressions needed to compute observables for the process $\mathrm{p}\,\mathrm{p} \to V + h$ ($V=$ $Z$, $W$) at NLO accuracy in the antenna subtraction formalism.
The formulae are given in a form valid both for $Z$ and $W$ production as indicated by the label $V$ appearing in the matrix element and phase space.
Before discussing the individual contributions, we first provide some general comments on our notation of the matrix elements and the overall factors used in the remainder of this section.
\begin{itemize}
\item
The matrix element denoted as $B^{\ell}_{n}$ corresponds to processes involving one correlated $q\bar{q}$ pair, $n$ gluons and a vector boson at $\ell$ loops. At NLO, we will also require the sub-leading colour matrix element $\tilde B^{\ell}_{n}$, the $N_F$ part $\hat B^{\ell}_{n}$, as well as $C^{\ell}_{n}$ including two $q\bar{q}$ pairs and $D^{\ell}_{n}$, which is the interference part of the processes with two $q\bar{q}$ pairs of identical flavour.
\item
Although we employ a non-diagonal CKM matrix in our computation for the process involving a $W$ boson, we refrain from explicitly including the flavour assignment and CKM dependence for clarity and generality between the neutral- and charged-current processes.
This information can easily be reinstated from the given formulae if necessary.
\item
For $V=W$, the initial state denoted as $q\bar q$ corresponds to $q \bar q''$, with $q''$ being the isospin partner of $q$. We assume that the flavour of $q$ ($\bar q$) is fixed (or becomes $\bar q''$ ($q''$) if a $W$ is coupled) while new quark flavours that appear at NLO are denoted as $\qp$ ($\bar q'$), also admitting $q = \qp$ ($\bar q = \bar q '$).
\item
The overall colour-independent factors that are denoted as $N_{ij}$ for $i,j=q,\bar{q},g$ below depend on the initial-state configuration and are defined as
\begin{eqnarray}
\label{eq:factors}
N_{qg}=N_{\bar q g}=\frac{g_s^2\mathcal{C}_V^2(N^2-1)}{4N(N^2-1)} \;\;\text{and} \;\; N_{q\bar q}=N_{q\bar q'}=N_{q q}=N_{q q'}=\frac{g_s^2\mathcal{C}_V^2(N^2-1)}{[2N]^2},
\end{eqnarray}
where $g_s^2= 4\pi\alpha_s$ and $\mathcal{C}_V=2(4\pi \alpha_V(M_V))$ are the QCD and electroweak couplings respectively, where the latter differ for the $Z$ and $W$ cases~\cite{Daleo:2006xa}: 
\begin{equation}
\alpha_{W}=\frac{G_F\,M_W^2\sqrt{2}}{4\pi}\,,\qquad 
\alpha_{Z}=\frac{G_F\,M_Z^2\sqrt{2}}{64\pi}\,.
\end{equation}
\item
For ease of readability, we fix the position of the identified parton to be 3 with an additional superscript (id.), while the initial-state particles are assigned to positions 1 and 2.
\end{itemize}

\subsection{Leading order}
\label{subsec:sig_lo}
At leading order, the short-distance cross section for $V +h$ receives contributions from three different sub-processes with initial states, $qg$, $\bar q g$ and $q\bar q$:
\begin{eqnarray}
	\der\hsig^{\rm{LO}}=\der\hsig^{\rm{LO}}_{(qg)}+\der\hsig^{\rm{LO}}_{(\bar{q} g)} + \der\hsig^{\rm{LO}}_{(q\bar{q})}\,.
	\label{eq:lochannels}
\end{eqnarray}
They are in one-to-one correspondence to the case of an identified quark ($q$), anti-quark ($\bar{q}$), and gluon ($g$) and read
\begin{eqnarray}
\der\hsig_{q}^{{\rm LO}}=N_{qg}B^{0}_{1}(1_{q},2_{g},\id{3}_{{q}},4_{V})
\times	\der\Phi_{2}(k_{{V{}}},\id{k}_3;p_1,p_2) \JET_{1}^{(1)}(\{\id{k}_{3}\};\eta \id{k}_{3})\,,
\\
\der\hsig_{\bar{q}}^{{\rm LO}}=N_{\bar q g}B^{0}_{1}(\id{3}_{{\bar{q}}},2_{g},1_{\bar{q}},4_{V})
\times	\der\Phi_{2}(k_{{V{}}},\id{k}_3;p_1,p_2) \JET_{1}^{(1)}(\{\id{k}_{3}\};\eta \id{k}_{3})\,,
\\
\der\hsig_{g}^{{\rm LO}}=N_{q \bar q}B^{0}_{1}(1_{q},\id{3}_{g},2_{\bar{q}},4_{V})
\times	\der\Phi_{2}(k_{{V{}}},\id{k}_3;p_1,p_2) \JET_{1}^{(1)}(\{\id{k}_{3}\};\eta \id{k}_{3})\,.
\end{eqnarray}

\subsection{Next-to-leading order}
\label{subsec:3_nlo}
In this section, we explicitly construct the NLO corrections to all sub-processes and possible cases of identified partons.

\subsubsection{Real matrix element contributions}
\label{subsubsec:3_real}
The real corrections receive contributions from the following initial-state parton-level processes:
\begin{eqnarray}
	\der\hsig^{\R} = \der\hsig_{(qg)}^{\R}+\der\hsig_{(\bar qg)}^{\R}+\der\hsig_{(q\bar q)}^{\R}+\der\hsig_{(gg)}^{\R}+\der\hsig_{(qq')}^{\R}+\der\hsig_{(q\bar q')}^{\R}+\der\hsig_{(\bar q \bar q')}^{\R}\,,
\end{eqnarray}
which we decompose into separate identified-parton pieces.
The case in which a quark $q$ is identified has contributions from the channels 
$qg$, $gg$, $qq'$ and $q \bar q'$:
\begin{eqnarray}
\der\hsig_{q}^{\R} &=&\alpha_s   \frac{\bar C(\epsilon)}{ C(\epsilon)}  \Big(  \mathcal{M}_{q,(qg)}^{\R}+\mathcal{M}_{q,(gg)}^{\R}+\mathcal{M}_{q,(qq')}^{\R} +\mathcal{M}_{q,(q\bar q')}^{\R}\Big) 
\nonumber \\ &&
\times \JET_{1}^{(2)}(\{\id{k}_{3},k_j\};\eta \id{k}_3)\,  \der\Phi_{3}(k_V,\id{k}_{3},k_j;p_1,p_2) \,.
\label{eq:sigRq}
\end{eqnarray}
The corresponding matrix elements denoted as $\mathcal{M}_{q}$ can be further colour decomposed according to the colour-ordered matrix elements introduced above:
\begin{eqnarray}
\mathcal{M}_{q,(qg)}^{\R}&=&  N_{qg}  N 
\Big\{ B_{2}^0 (1_q,2_g,j_g,\id{3}_q,5_V) 
+ [2_g \leftrightarrow j_g] 
-\frac{1}{N^2}
 \tilde  B_{2}^0 (1_q,2_g,j_g,\id{3}_q,5_V) \Big\} \,, \;           
\label{eq:sigRqqg} \\
\mathcal{M}_{q,(gg)}^{\R} &=&  N_{gg}   N \sum_{i=u,d}  N_i\Big\{  B_{2}^0 (j_{\bar{i}},1_g,2_g,\id{3}_i,5_V) + [1_g \leftrightarrow 2_g]
\nonumber \\ &&
-\frac{1}{N^2} \tilde B_{2}^0 (j_{\bar{i}},1_g,2_g,\id{3}_i,5_V)  \Big\}   \,, \;\;
\label{eq:sigRqgg}\\
\mathcal{M}_{q,(q q')}^{\R} &=& 
 N_{qq'}  \Big\{ C_{0}^0 (1_q,j_{q'},2_{q'},\id{3}_{ q},5_V) 
  - \frac{1}{N} {D} _0^0( 1 _q, j_{ q'},2_{q'},\id 3 _{ q},5_V) \Big\} \,,
\label{eq:sigRqqqp}\\
\mathcal{M}_{q,(q \bar q')}^{\R} &=&   N_{q\bar q'}  C_{0}^0 (1_q,2_{\bar q'},j_{\bar q'},\id{3}_{ q},5_V)   \,.
\label{eq:sigRqqqbp}
\end{eqnarray}
Note that in practice, all $N_F$ contributions are split into $N_u$ and $N_d$ types due to the different electroweak couplings of the $u$- and $d$-type quarks to the vector boson.
Further, the identification of $q$ and $q'$ has to be performed independently since their flavour correlations are different. 
\\
When $q'$ is the identified parton, the real level short-distance contributions read: 
\begin{eqnarray}
	\der\hsig_{q'}^{\R} &=& \alpha_s \frac{\bar C(\epsilon)}{ C(\epsilon)} \Big(\mathcal{M}_{q',(q q')}^{\R}+ \mathcal{M}_{q',(q \bar q)}^{\R}  
	\Big)
	\nonumber \\ &&
	\times \JET_{1}^{(2)}(\{\id{k}_{3},k_j\};\eta \id{k}_3) \,  \der\Phi_{3}(k_V,\id{k}_{3},k_j;p_1,p_2) \,,
	\label{eq:sigRqpid}
\end{eqnarray}
with
\begin{eqnarray}
	\mathcal{M}_{q',(q q')}^{\R}&=&  N_{qq'} \Big\{C_0^0 (1_q,\id{3}_{q'},2_{q'},j_{ q},5_V) 
	- \frac{1}{N} {D} ^0_0( 1 _q,\id 3 _{ q'},2_{q'}, j_{ q},5_V) \Big\} \,,
	\label{eq:sigRqp}\\
	\mathcal{M}_{q',(q \bar q)}^{\R} &=& 
	N_{q\bar q}   \Big\{ \sum_{i=u,d} N_i C_{0}^0 (1_q,\id{3}_{i},j_{\bar i},2_{\bar q},5_V) - \frac{1}{N} {D} ^0_0( 1 _q,\id 3 _{\bar q},j_q, 2_{\bar q},5_V)  \Big\}   \,,
	\label{eq:sigRqqqb}
\end{eqnarray}
where eq.~\eqref{eq:sigRqp} does not contain any singular limits. The contributions when the anti-quark $\bar{q}$ or $\bar{q}'$ are identified are respectively similar to eqs.~\eqref{eq:sigRq}--\eqref{eq:sigRqqqbp} or eqs.~\eqref{eq:sigRqpid}--\eqref{eq:sigRqqqb} with the exchange of quarks and anti-quarks in both initial and final states.
\\
We focus next on the real-level partonic contributions where a gluon $g$ is the identified particle.
Those contributions arise from the $qg$, $\bar qg$ and $q \bar q$ initial-state combinations and read:
\begin{eqnarray}
\der\hsig_{g}^{\R} &=& \alpha_s \frac{\bar C(\epsilon)}{ C(\epsilon)} \Big( \mathcal{M}_{g,(q g)}^{\R} +\mathcal{M}_{g,(\bar q g)}^{\R} +\mathcal{M}_{g,(q \bar q)}^{\R}
   \Big)
\nonumber \\ &&
 \times \JET_{1}^{(2)}(\{\id{k}_{3},k_j\};\eta \id{k}_3) \,  \der\Phi_{3}(k_V,\id{k}_{3},k_j;p_1,p_2) \,,
\label{eq:sigRg}
\end{eqnarray}
where the contributions from $\mathcal{M}_{g,(q g)}^{\R}$ and $\mathcal{M}_{g,(\bar q g)}^{\R}$ are the same upon the exchange $q \leftrightarrow \bar q$. Explicitly, the matrix elements labelled as $\mathcal{M}_{g}$ from the channels
$qg$ and $q \bar{q}$ admit the following colour decomposition:
\begin{eqnarray}
\mathcal{M}_{g,(q g)}^{\R} &=&
N_{qg}  N \Big\{
\,B_{2}^0 (1_q,2_g,\id{3}_g,j_q,5_V)
+ [\id{3}_{g}\leftrightarrow 2_g]
\nonumber \\ &&
- \, \frac{1}{N^2} 
\tilde B_{2}^0 (1_q,2_g,\id{3}_g,j_q,5_V) \Big\} \,,
\label{eq:sigRgqg}\\
\mathcal{M}_{g,(q\bar q)}^{\R} &=& 
 N_{q\bar q} \frac{N}{2!}
 \Big\{  B_{2}^0 (1_q,\id{3}_{g},j_{g},2_{\bar q},5_V)+ [\id{3}_{g}\leftrightarrow j_g] 
 \nonumber \\ &&
-  \frac{1}{N^2} \tilde B_{2}^0 (1_q,\id{3}_{g},j_{g},2_{\bar q},5_V)  \Big\}  \,.
\label{eq:sigRgqqb}
\end{eqnarray}
%
\subsubsection{Real subtraction terms}
\label{subsubsec:3_real_sub}
The real-level subtraction terms are made of two categories of terms depending if the unresolved parton $j$ is colour connected to the identified parton or not.
If the fragmenting particle is not colour connected to the unresolved parton, the subtraction terms mirror those
for $V+\rm{jet}$ production~\cite{Daleo:2006xa}. These terms are associated to initial--initial kinematical configurations for which we use the notation $\widetilde{1}_p$ and $\widetilde{2}_p$ and no <<id.>> index on the $X^3_0$ antenna functions.
When instead, the fragmenting particle participates actively in the unresolved limit, an initial--final fragmentation subtraction of the type given in eq.~\eqref{eq:SNLOxs} is needed. These terms include the fragmentation antennae denoted with the <<id.>> index and we use $\widetilde{1}_p$ or $\widetilde{2}_p$, together with $\id{\widetilde{(3j)}}_p$ to indicate the mapped momenta. These subtraction terms have been constructed for the first time for the computations presented in this work.
For each of the identified particles, we present the real-level subtraction terms including both categories.

The real-level subtraction terms associated to parton $j$ becoming unresolved and corresponding to an identified quark $q$  denoted as
$\der\hsig_{q}^{\S}$ has itself four separate contributions related to the initial states $qg$, $gg$, $qq'$ and $q\bar{q}'$. 
The sum of those contributions takes the form: 
\begin{eqnarray}
\der\hsig_{q}^{\S}&=& \alpha_s \frac{\bar C(\epsilon)}{ C(\epsilon)}  \Big( \mathcal{M}_{q(q),(qg)}^{\S}+\mathcal{M}_{q(q),(gg)}^{\S}+\mathcal{M}_{q(q),(qq')}^{\S}+\mathcal{M}_{q(q),(q\bar q')}^{\S}\, \Big)
\nonumber \\ &&
\times {J}_{1}^{(1)}(\{\id{{k}}_3 \};\eta \id{{k}}_3)  \der\Phi_{3}(k_V,\id{k}_{3},k_j;p_1,p_2)\,.
\label{eq:sigSqq}
\end{eqnarray}
The explicit expressions for the contributions from each initial-state channel read:
\begin{eqnarray}
\mathcal{M}_{q(q),(qg)}^{\S}&=& 
N_{qg} N \Big\{  
 \, D_3^0(1_q,j_g,2_g)\, 
B_1^0(\widetilde{1}_q,\widetilde{2}_g,\id{3}_q,4_V)
\nonumber \\ && \phantom{\times \times }
+ \, d_3 ^{0, \text{id}. q}(\id{3}_q,j_g,2_g)\, 
B_1^0({1}_q,\widetilde{2}_g,\id{\widetilde{(3j)}}_q,4_V)\, 
\nonumber \\ &&  \phantom{\times \times }
- \frac{1}{N^2}\,  A_3^{0, \text{id}. q}(\id{3}_q,j_g,1_{ q}) \,
B_1^0(\widetilde{1}_q,{2}_g,\id{\widetilde{(3j)}}_q,4_V) \,
 \Big\}                                       ,
\label{eq:sigSqqg}\\
\mathcal{M}_{q(q),(gg)}^{\S}&=& 
-N_{gg}  N_F N 
\Big( 1 + \frac{1}{N^2} \Big)\Big( d_3^0(j_{\bar q},2_g,1_g)
B_1^0(\id{3}_q,\widetilde{1}_g,\widetilde{2}_q,4_V))
-[\widetilde{1}_q \leftrightarrow \widetilde{2}_g] \Big)\,, \;\;\;\;\;\;\;
\label{eq:sigSqgg} 
\\
\mathcal{M}_{q(q),(q\bar q')}^{\S} &\stackrel{\bar q'  \leftrightarrow  q'}{=}& \mathcal{M}_{q(q),(qq')}^{\S} = 
N_{gq'}  E_3^0({1}_{ q},j_{\bar q'},2_{ q'})\, 
B_1^0(\widetilde{1}_q,\widetilde{2}_g,\id{3}_q,4_V)  \,.
\label{eq:sigSqqqp}
\end{eqnarray}

The contribution where the quark $q'$ is identified has only a contribution from the initial state $q\bar{q}$. It reads:
\begin{eqnarray}
\der\hsig_{q'}^{\S} &=& \alpha_s \frac{\bar C(\epsilon)}{ C(\epsilon)}  \Big(   \mathcal{M}_{g(q'),(q\bar q)}^{{\S}} \Big) 
 {J}_{1}^{(1)}(\{\id{{k}}_3 \};\eta \id{{k}}_3)  \der\Phi_{3}(k_V,\id{k}_{3},k_j;p_1,p_2)\,.
\end{eqnarray}
%
This equation is related to a contribution in which the flavour of the identified particle in the reduced matrix element is not the same as the one in the real correction. This is usually denoted as an identity changing (IC) subtraction term. In this specific case, the quark $q'$ becomes a gluon $g$ in the reduced matrix element.
Further expanding in products of antennae and reduced matrix elements, the individual contribution read:
\begin{eqnarray}
\mathcal{M}_{g(q'),(q\bar q)}^{{\S}}  &=& 
N_{q\bar q} 
 N_F  E_3^{0, \text{id}. q}(1_{q},{j}_{\bar q'},\id3_{ q'})\, 
B_1^0(\widetilde{1}_{q},\id{\widetilde{(3j)}}_g,2_{\bar q},4_V)\,, 
\label{eq:sigSqqb_qpIC}
\end{eqnarray}
where the flavour changing feature is specified with the labelling $g(q')$ on the left-hand side of the equation.
\\
Finally, we present the explicit expression for the subtraction term involving the identified parton to be the gluon.
The general expression is the sum of three terms:
\begin{equation}
  \der\hsig_{g}^{\S} =
  \der\hsig_{g(g)}^{\S} + \der\hsig_{q(g)}^{\S} + \der\hsig_{\qb(g)}^{\S}\,.
  \label{eq:sigSg}
\end{equation}
The first term denoted as $ \der\hsig_{g(g)}^{\S}$ is of identity preserving nature and reads: 
\begin{eqnarray}
\der\hsig_{g(g)}^{\S}&=&  \alpha_s \frac{\bar C(\epsilon)}{ C(\epsilon)}  \Big( \mathcal{M}_{g(g),(qg)}^{{\S}}+\mathcal{M}_{g(g),(\bar qg)}^{{\S}}+\mathcal{M}_{g(g),(q\bar q)}^{{\S}} \Big)
\nonumber \\ &&
\times {J}_{1}^{(1)}(\{\id{{k}}_3 \};\eta \id{{k}}_3)  \der\Phi_{3}(k_V,\id{k}_{3},k_j;p_1,p_2)\,,  
\label{eq:sigSgg}
\end{eqnarray}
with
\begin{eqnarray}
\mathcal{M}_{g(g),(\bar qg)}^{\S} &\stackrel{q  \leftrightarrow \bar q}{=}& \mathcal{M}_{g(g),(qg)}^{\S}= -N_{qg} N 
\Big\{ \Big( 1-\frac{1}{N^2} \Big)  A_{3}^0(1_q,2_g,j_q)\, B_1^0(\widetilde{1}_{{q}},\id{3}_g,\widetilde{2}_{{\bar{q}}},4_V) \Big\} \,, \;\;\;
\label{eq:sigSggqg}\\
\mathcal{M}_{g(g),(q\bar q)}^{\S}&=& 
N_{q\bar q} N \Big\{ 
\Big(d_{3}^{0, \text{id}. g}(1_{{q}},\id{3}_{g},j_g) + [j_{{g}} \leftrightarrow \id{3}_{g}]\Big)\, 
B_1^0(\widetilde{1}_{q},\id{\widetilde{({3}j)}}_{g},2_{\bar q},4_V)\, 
\nonumber \\ && \phantom{ \times  \times}
+\Big( d_{3}^{0, \text{id}. g}(2_{\bar{q}},i_g,\id{3}_{g})+ [j_g  \leftrightarrow \id{3}_{g}] \Big) \, 
B_1^0({1}_{q},\id{\widetilde{({3}j)}}_{g},\widetilde{2}_{\bar q},4_V)\, 
\nonumber \\ && \phantom{ \times  \times}
+\frac{1}{N^2} \, A_{3}^0(1_q,j_g,2_{\bar q})\, 
B_1^0(\widetilde{1}_{{{q}}},\id{3}_g,\widetilde{2}_{{\bar q}},4_V)
\Big\}\,.  
\label{eq:sigSggqqb}
\end{eqnarray}
%
The second and third terms of eq.~\eqref{eq:sigSg} are of identity changing nature and we have $\der\hsig_{\bar q(g)}^{\S} \stackrel{q  \leftrightarrow \bar q}{=}  \der\hsig_{q(g)}^{\S}$. In this case, only the $qg$  ($\bar q g$) channel contributes:
\begin{eqnarray}
\der\hsig_{q(g)}^{\S} &=&  \alpha_s \frac{\bar C(\epsilon)}{ C(\epsilon)} \Big( \mathcal{M}_{q(g),(qg)}^{\S} \Big) {J}_{1}^{(1)}(\{\id{{k}}_3 \};\eta \id{{k}}_3)  \der\Phi_{3}(k_V,\id{k}_{3},k_j;p_1,p_2) \,,
\end{eqnarray}
with
\begin{eqnarray}
\mathcal{M}_{q(g),(qg)}^{\S}&=&N_{qg} N \Big\{ 
 d_{3}^{0, \text{id}. g}(j_q,\id{3}_g,2_g)\, 
B_1^0({1}_q,\widetilde{2}_{g},\id{\widetilde{({3}j)}}_q,4_V)
\nonumber \\ && \phantom{\times \times}
+\frac{1}{N}  \, A_{3}^{0, \text{id}. g}(1_q,\id{3}_g,j_q)\, 
B_1^0(\widetilde{1}_{q},2_g,\id{\widetilde{({3}j)}}_{ q},4_V) \Big\} \,. 
\label{eq:sigSgg2}
\end{eqnarray}
\subsubsection{Virtual matrix element contributions}
\label{subsubsec:3_virt}
The virtual one-loop contributions arise from the same channels as the ones present at Born level and given in eq.~\eqref{eq:lochannels}. Those comprise three initial-state configurations that are associated with the identified partons being a quark, an anti-quark or a gluon,
\begin{eqnarray}
	\der\hsig^{\V}_{} = \alpha _s \bar C(\epsilon)\Big( \mathcal{M}^{\V}_{q,(qg)}+\mathcal{M}^{\V}_{\bar{q},(\bar q g)}+\mathcal{M}^{\V}_{g,(q\bar q)} \Big)\JET_{1}^{(1)}(\{\id{k}_{3}\};\eta \id{k}_{3})  \der\Phi_{2}(k_{V},\id{k}_3;p_1,p_2)\,, \;\;\;\;
\end{eqnarray}
where,
\begin{eqnarray}
\mathcal{M}^{\V}_{q,(qg)} &=& 
N_{qg}  \, N \Big\{  B^{1}_{1}(1_{q},2_{g},\id{3}_{{q}},4_V)
- \frac{1}{N^2} \tilde B^{1}_{1}(1_{q},2_{g},\id{3}_{{q}},4_V)
\nonumber \\ &&
 + \frac{N_F}{N} \hat B^{1}_{1}(1_{q},2_{g},\id{3}_{{q}},4_V)\, \Big\} \,,
 \label{eq:sigVq} \\
\mathcal{M}^{\V}_{g,(q\bar q)} &=& 
N_{q\bar q}
 N \Big\{   B^{1}_{1}(1_{q},\id{3}_{g},2_{\bar{q}},4_V)
- \frac{1}{N^2} \tilde  B^{1}_{1}(1_{q},\id{3}_{g},2_{\bar{q}},4_V)
\nonumber \\ &&
 + \frac{N_F}{N} \hat  B^{1}_{1}(1_{q},\id{3}_{g},2_{\bar{q}},4_V)\, \Big\} \,, 
 \label{eq:sigVg}
\end{eqnarray}
with $\mathcal{M}^{\V}_{q,(\bar qg)} \stackrel{q  \leftrightarrow \bar q}{=}\mathcal{M}^{\V}_{q,(qg)}$.

The infrared behaviour of the virtual matrix elements $B^1_1$, $\tilde B^1_1$ and $\hat B^1_1$ is the same regardless of the type of vector boson and can be expressed in terms of the Catani one-loop factorisation formula~\cite{Catani:1998bh} with the colour-ordered singularity operators $\textbf{I}_{ij}^{(1)}$ whose forms are given explicitly in the appendix of~\cite{Currie:2013vh}. For the process at hand, we have three different crossings for the $B$-type matrix element, one for each channel. In the case of the $qg$-initiated process, the pole structure is
\begin{eqnarray}
  {\rm Poles}\Big(B^{1}_{1}(1_{q},2_{g},{3}_{{q}},4_V)\Big) &=& 2 \Big( \mathbf{I}^{(1)}_{q g}(\e,s_{12})
  +  \mathbf{I}^{(1)}_{qg}(\e,s_{23}) \Big)
   \times B^{0}_{1}(1_{q},2_{g},{3}_{{q}},4_V)\,, \label{eq:polesB11qqbLC} \\ 
 {\rm Poles}\Big(\tilde B^{1}_{1}(1_{q},2_{g},{3}_{{q}},4_V)\Big) &=& 2 \Big( \mathbf{I}^{(1)}_{qq}(\e,s_{13})\Big) 
   \times B^{0}_{1}(1_{q},2_{g},{3}_{{q}},4_V)\,,\\
 {\rm Poles}\Big(\hat B^{1}_{1}(1_{q},2_{g},{3}_{{q}},4_V)\Big) &=& 2 \Big( \mathbf{I}^{(1)}_{qg,F}(\e,s_{12}) + \mathbf{I}^{(1)}_{qg,F}(\e,s_{23})\Big)
   \times B^{0}_{1}(1_{q},2_{g},{3}_{{q}},4_V)\,,
\label{eq:polesB11qg}
\end{eqnarray}
which is the same as for the $\bar q g$-channel. For $q \bar q$-initiated process and in the case of $Z+h$ or for the $q \bar q''$ initiated process in the case of $W+h$ production, the pole structure of the virtual matrix element contributions
take the form:
\begin{eqnarray}
  {\rm Poles}\Big(B^{1}_{1}(1_{q},{3}_{g},2_{\bar{q}},4_V)\Big) &=& 2 \Big( \mathbf{I}^{(1)}_{qg}(\e,s_{13})
  +  \mathbf{I}^{(1)}_{qg}(\e,s_{23}) \Big)
   \times B^{0}_{1}(1_{q},{3}_{g},2_{\bar{q}},4_V)\,, \\
 {\rm Poles}\Big(\tilde B^{1}_{1}(1_{q},{3}_{g},2_{\bar{q}},4_V)\Big) &=& 2 \Big( \mathbf{I}^{(1)}_{q {q}}(\e,s_{12}) \Big)
   \times B^{0}_{1}(1_{q},{3}_{g},2_{\bar{q}},4_V)\,, \\
 {\rm Poles}\Big(\hat B^{1}_{1}(1_{q},3_{g},2_{\bar{q}},4_V)\Big) &=& 2 \Big( \mathbf{I}^{(1)}_{qg,F}(\e,s_{13})
  +  \mathbf{I}^{(1)}_{qg,F}(\e,s_{23}) \Big)
   \times B^{0}_{1}(1_{q},{3}_{g},2_{\bar{q}},4_V)\,.
\label{eq:polesB11qqb}
\end{eqnarray}
\subsubsection{Virtual subtraction  terms}
\label{subsubsec:3_virt_sub}
As discussed in section~\ref{subsubsec:theorysection_virtual}, we construct the virtual-level subtraction term as
 \begin{equation}
	\der\hsig^\T(\eta,x_1,x_2,z) =  - \int _1 \der\hsig^{\S}(\eta,x_1,x_2,z) -\der\hsig^{\MF}(\eta,x_1,x_2,z)  \,,
\end{equation}
where $\int _1 \der\hsig^{\S}$ is the integrated real-level subtraction term and $\der\hsig^{\MF}$ the mass-factorisation term. We combine both terms using integrated fragmentation dipoles  $\boldsymbol{J}_{2} ^{(1),\text{id.}p} $ as defined
in eq.~\eqref{eq:J21def} as well as the standard integrated dipoles $\boldsymbol{J}_{2} ^{(1)} $~\cite{Currie:2013vh}.


We present the complete virtual subtraction term separately for the different identified partons. For the identified-quark ($q$) contribution, the virtual subtraction term denoted as $\der\hsig_{q}^{\T}$ reads:
\begin{eqnarray}
	\der\hsig_{q}^{\T}&=& \int \frac{\der x_1}{x_1} \frac{\der x_2}{x_2} \der z\, \alpha_s {\bar C(\epsilon)}  \big( \mathcal{M}_{q(q),(qg)}^{\T}+\mathcal{M}_{q(q),(gg)}^{\T}+\mathcal{M}_{q(q),(qq')}^{\T}+\mathcal{M}_{q(q),(q\bar q')}^{\T}\, \big)
	\nonumber \\ && \times
{J}_{1}^{(1)}(\{\id{k}_3 \};\eta \id{k}_3)  \der\Phi_{2}(k_V,\id{k}_{3};p_1,p_2) \,,
	\label{eq:sigTqid}
\end{eqnarray}
where 
\begin{eqnarray}
	\mathcal{M}_{q(q),(qg)}^{\T}&=& 
N_{qg}N \big\{  
\, \big(  \boldsymbol{J} _{2,QG}^{(1),II}(s_{12})\, 
+ \, \boldsymbol{J} _{2,QG} ^{(1),FI, \text{id}. q}(s_{23})\, 
- \frac{1}{N^2}\, \boldsymbol{J} _{2,QQ} ^{(1),IF, \text{id}.q }(s_{13}) \nonumber \\ && \phantom{\times \times }
-2 N_F\boldsymbol{J} _{2,hQG}^{(1),II}(s_{12})\big) 
\times B_1^0({1}_q,{2}_g,\id{3}_q,4_{V})\, \big\}
{J}_{1}^{(1)}(\{\id{k}_3 \};\eta \id{k}_3)  \,, \;
\label{eq:sigTqq}
\\	
	\mathcal{M}_{q(q),(gg)}^{\T}&=&  N_{gg}N_F \big(N-{1}/{N}\big) \big\{ 
-\boldsymbol{J} _{2,GQ,g\rightarrow q} ^{(1),II }(s_{12})B_1^0({1}_q,{2}_g,\id{3}_q,4_V)  
\nonumber \\ && \phantom{\times \times \times \times \times \times \; \; \;  \times  } 
-\boldsymbol{J} _{2,QG,g\rightarrow q} ^{(1),II }(s_{12})B_1^0({2}_q,{1}_g,\id{3}_q,4_V) \big\} \,,
\\
	\mathcal{M}_{q(q),(q\bar q')}^{\T}&\stackrel{q  \leftrightarrow \bar q}{=}&\mathcal{M}_{q(q),(qq')}^{\T}= - N_{qq'}  \boldsymbol{J} _{2,QG,q'\rightarrow g} ^{(1),II }(s_{12})  \, 
	B_1^0({1}_q,{2}_g,\id{3}_q,4_V)\,.
	\label{eq:sigTqqp}
\end{eqnarray}

In the case where the quark $q'$ is the identified particle, the virtual subtraction term $\der\hsig_{q'}^{\T}$ reads:
\begin{eqnarray}
	\der\hsig_{q'}^{\T} &=& \int \frac{\der x_1}{x_1} \frac{\der x_2}{x_2} \der z\,\alpha_s {\bar C(\epsilon)}  \Big( \mathcal{M}_{g(q'),(q\bar q)}^{{\T}} \Big) 
{J}_{1}^{(1)}(\{\id{k}_3 \};\eta \id{k}_3)  \der\Phi_{2}(k_V,\id{k}_{3};p_1,p_2) \,,
\label{eq:sigTqpid}
\end{eqnarray}
%
with
\begin{eqnarray}
	\mathcal{M}_{g(q'),(q\bar q)}^{\T}  &=&   -N_{qg}	N_F   \boldsymbol{J} _{2,QG,q\leftarrow g} ^{(1),IF, \text{id}.q' }(s_{13}) 
B_1^0({1}_{q},\id{{3}}_g,2_{\bar q,},4_V)\,.
\end{eqnarray}
For the case where $\bar q$ and $\bar q'$ are the identified partons, the terms have the same structure as for $q$ and $q'$ respectively. Finally, the case in which the gluon is the identified parton is given by:
\begin{eqnarray}
\der\hsig_{g}^{\T}&=&\int \frac{\der x_1}{x_1} \frac{\der x_2}{x_2} \der z\,  \alpha_s \frac{\bar C(\epsilon)}{ C(\epsilon)}  \Big( \mathcal{M}_{g(g),(qg)}^{{\T}}+\mathcal{M}_{g(g),(\bar qg)}^{{\T}}+\mathcal{M}_{g(g),(q\bar q)}^{{\T}}+ \mathcal{M}_{q(g),(qg)}^{\T} 
\nonumber \\ && 
 + \mathcal{M}_{q(g),(\bar qg)}^{\T} \Big) \times
{J}_{1}^{(1)}(\{\id{k}_3 \};\eta \id{k}_3)  \der\Phi_{2}(k_V,\id{k}_{3};p_1,p_2) \,,
\label{eq:sigTgid}
\end{eqnarray}
where
\begin{eqnarray}
\mathcal{M}_{g(g),(qg)}^{\T}&=& - N_{qg} \Big( N-	\frac{1}{N}\Big) 
\boldsymbol{J} _{2,QQ,g\to q} ^{(1),II }(s_{12}) 
B_1^0({1}_{{q}},\id{3}_g,{2}_{{\bar{q}}},4_V) 
\,,  
\label{eq:sigTggqg}\\
	\mathcal{M}_{g(g),(q\bar q)}^{\T}&=&   -N_{qg} N 
\Big( \boldsymbol{J} _{2,GQ} ^{(1),FI, \text{id}.g }(s_{23}) + \boldsymbol{J} _{2,QG} ^{(1),IF, \text{id}.g }(s_{13}) + \boldsymbol{J} _{2,QQ} ^{(1),II }(s_{12})
\nonumber \\ &&
 +N_F  \big( \boldsymbol{J} _{2,QG,q\leftarrow g} ^{(1),IF, \text{id}.q' }(s_{13})  +\boldsymbol{J}_{2,hQG}^{(1),FI, \text{id}.g }(s_{13}) \big)
\Big) B_1^0({1}_{{{q}}},\id{3}_g,{2}_{{\bar q}},4_V)\,,  
\label{eq:intTggqqb} \\
\mathcal{M}_{q(g),(qg)}^{\T}&=&  N_{qg}  N 
\Big( -\boldsymbol{J} _{2,QG,g \leftarrow q} ^{(1),FI, \text{id}.g }(s_{23}) 
\nonumber \\ &&
-\frac{1}{N^2}  \boldsymbol{J} _{2,QQ,g \leftarrow q} ^{(1),IF, \text{id}.g }(s_{13}) \Big) 
B_1^0({1}_{q},2_g,\id{{{3}}}_{ q},4_V) \,,
\label{eq:intsigTgg4}
\end{eqnarray}
with the contributions from the $\bar qg$ channel being the same as the $qg$ channel after the exchange of the quark for an anti-quark.

The virtual subtraction dipoles denoted as $\boldsymbol{J}_{2}^{(1)}$ can be separated into two categories depending whether there is (or not) a colour connection between the identified particle and the unresolved parton.
In the case where there is no colour connection, the virtual subtraction dipoles are all of the initial--initial type.
Those are known~\cite{Currie:2013vh} and recalled here for completeness:
\begin{eqnarray}
  \boldsymbol{J}_{2,QG}^{1,II} &=& \delta(1-z) \left[ \mathcal D_{3,qg}^0(x_1,x_2) - \delta(1-x_2)\Gamma^{(1)}_{qq}(x_1)
    - \frac{1}{2} \delta(1-x_1)\Gamma^{(1)}_{gg}(x_2) \right]\,,\\
\boldsymbol{J}_{2,QQ} ^{(1),II} &=& \delta(1-z) \left[ \mathcal A_{3,q\bar{q}}^0(x_1,x_2)-\delta(1-x_2)\Gamma^{(1)}_{qq}(x_1) -\delta(1-x_1)\Gamma^{(1)}_{qq}(x_2) \right]\,,\\	
\boldsymbol{J}_{2,QQ,g\to q}^{(1),II}&=& \delta(1-z) \left[ - \mathcal A_{3,qg}^0(x_1,x_2) - \delta(1-x_1)\Gamma^{(1)}_{qg}(x_2) \right]\,,\\
\boldsymbol{J}_{2,GQ,g\rightarrow q}^{(1),II}&=& \delta(1-z) \left[ - \mathcal D_{3,gg}^0(x_1,x_2) - \delta(1-x_2)\Gamma^{(1)}_{qg}(x_1) \right]\,,\\
\boldsymbol{J}_{2,QG,g\rightarrow q}^{(1),II}&=& \delta(1-z) \left[ - \mathcal D_{3,gg}^0(x_1,x_2) - \delta(1-x_1)\Gamma^{(1)}_{qg}(x_2) \right]\,,\\
\boldsymbol{J}_{2,QG,q'\rightarrow g}^{(1),II}&=& \delta(1-z) \left[ - \mathcal E_{3,q'q}^0(x_1,x_2) - \delta(1-x_2)\Gamma^{(1)}_{gq}(x_1) \right]\,,\\
\boldsymbol{J}_{2,hQG}^{(1),II}&=& -\frac{1}{2}\delta(1-z)\left[ \delta(1-x_1)\Gamma^{(1)}_{gg,F}(x_2) \right] \,.
\end{eqnarray}

For the case where the identified particle participates in the unresolved limit, the fragmentation
dipoles are derived here for the first time. Those are separated according to the nature of the identified parton involved.
In the case where a quark is identified, the required dipoles read:
\begin{eqnarray}
  \boldsymbol{J} _{2,QG} ^{(1),FI, \text{id}.q } &=&  \delta(1-x_1) \left[ \mathcal D_{3,g}^{0, \text{id}. q}(x_2,z)
    - \delta(1-x_2)\Gamma^{(1)}_{qq}(z) - {\frac{1}{2}}\delta(1-z)\Gamma^{(1)}_{gg}(x_2) \right]\,, \;\;\;\;\;\;\;\;\; \\
  \boldsymbol{J} _{2,QQ} ^{(1),IF, \text{id}.q } &=&  \delta(1-x_2) \left[ \mathcal A_{3,q}^{0, \text{id}. q}(x_1,z)
    - \delta(1-x_1)\Gamma^{(1)}_{qq}(z) - \delta(1-z)\Gamma^{(1)}_{qq}(x_1)\right]\,, \\
  \boldsymbol{J} _{2,QG, q \leftarrow g} ^{(1),IF, \text{id}.q' } &=& \delta(1-x_2) \left[ - \mathcal E_{3,q}^{0, \text{id}. q'}(x_1,z)
 +\delta(1-x_1) \Gamma^{(1)}_{qg}(z) \right] \,.
\end{eqnarray}
For the identified-gluon case, the dipoles used here instead read:
\begin{eqnarray}
  \boldsymbol{J} _{2,GQ} ^{(1),FI, \text{id}.g } &=&  \delta(1-x_1) \left[ \mathcal D_{3,q}^{0, \text{id}. g}(x_2,z)
    - \delta(1-z)\Gamma^{(1)}_{qq}(x_1) - {\frac{1}{2}}\delta(1-x_1)\Gamma^{(1)}_{gg}(z) \right]\,, \;\;\;\;\;\;\;\;\; \\
      \boldsymbol{J} _{2,QG} ^{(1),IF, \text{id}.g } &=&  \delta(1-x_2) \left[ \mathcal D_{3,q}^{0, \text{id}. g}(x_1,z)
    - \delta(1-z)\Gamma^{(1)}_{qq}(x_2) - {\frac{1}{2}}\delta(1-x_2)\Gamma^{(1)}_{gg}(z) \right]\,, \;\;\;\;\;\;\;\;\; \\
  \boldsymbol{J} _{2,QQ, g \leftarrow q} ^{(1),IF, \text{id}.g } &=& \delta(1-x_2) \left[ -\mathcal A_{3,q}^{0, \text{id}. g}(x_1,z)
   +\delta(1-x_1)\Gamma^{(1)}_{gq}(z) \right]\,,\\
  \boldsymbol{J} _{2,QG,g \leftarrow q} ^{(1),FI, \text{id}.g } &=& \delta(1-x_1) \left[- \mathcal D_{3,g} ^{0, \text{id}. g}(x_2,z)
   + \delta(1-x_2) \Gamma^{(1)}_{gq}(z) \right]\,, \\
   \boldsymbol{J}_{2,hQG}^{(1),FI, \text{id}.g }&=& -\frac{1}{2}\delta(1-x_2)\left[ \delta(1-x_1)\Gamma^{(1)}_{gg,F}(z) \right] \,.
\end{eqnarray}	
In the final-state identity-changing cases we use a right-to-left arrow, while in the initial-state identity-changing cases we use a left-to-right arrow.
We can sum all virtual subtraction terms presented in section~\ref{subsubsec:3_virt_sub}
and the corresponding virtual-level matrix element contributions presented in section~\ref{subsubsec:3_virt}
and show that it is free from explicit poles,
\begin{equation}\label{eq:subNLO2}
	Poles \bigg(  \der\hsig^{\V}(\eta) - \Big( \underbrace{- \int _1 \der\hsig^{\S}(\eta) -\der\hsig^{\MF}(\eta) }_{\equiv\der\hsig^{\T}(\eta)} \Big) \, \bigg) = 0 \,.
\end{equation} 
The finite remainder can then be integrated numerically.	

\section{\texorpdfstring{$Z$}{Z} boson in association with a hadron inside a jet}
\label{sec:Z+h}

In this section we provide predictions for observables related to the production of charged hadrons inside hadronic jets in association with a leptonically decaying $Z$-boson. We compare our results with the corresponding LHCb measurement~\cite{LHCb:2022rky} at $\sqrt{s} = 13$\ TeV and consider both unidentified charged hadron as well as pion production.

\subsection{Observable definition and setup}
\label{subsubsec:observable_lhcb}

Following~\cite{LHCb:2022rky}, the LHCb measurement is performed by considering events with at least one reconstructed hadronic jet and by identifying a hadron inside it.
In case multiple jets are reconstructed in the event, the one with the largest transverse momentum ($j_1$) is considered.
Determining the longitudinal momentum fraction carried by the hadron within the jet,
\begin{eqnarray}
\label{eq:z_observable_lhcb}
	z=\frac{\textbf{p}_{h}\cdot \textbf{p}_{j_1}}{|\textbf{p}_{j_1}|^2}\,,
\end{eqnarray}
where $\textbf{p}_{h}$ and $\textbf{p}_{j_1}$ are the hadron and the jet three-momenta respectively,
the measured observable is given by the normalised distribution
\begin{eqnarray}\label{eq:fz_experimental_definition}
	F_\text{exp.}(z)=\frac{1}{N_{Z+\rm{jet}}}\frac{\der N_{h}(z)}{\der z},
\end{eqnarray}
where $N_{h}(z)$ is the number of events with a hadron carrying a longitudinal momentum fraction $z$ inside the leading jet, while the normalization factor $N_{Z+\rm{jet}}$ corresponds to the number of $Z+\rm{jet}$ events.
On the theory side, eq.~\eqref{eq:fz_experimental_definition} is mirrored by
\begin{eqnarray}\label{eq:fz_theoretical_definition}
	F_\text{th.}(z)=\frac{1}{\sigma_{Z+\rm{jet}}}\frac{\der\sigma_{Z+h}}{\der z},
\end{eqnarray}
where the differential distribution present in the numerator require the presence of both a jet and a hadron inside it, whereas the inclusive cross section $\sigma_{Z+\rm{jet}}$ only demands the presence of a jet.

The LHCb measurement is performed within the following fiducial volume:
%
\begin{eqnarray}\label{eq:fiducialcuts_lhbc13}
	&20 <\ptj < 100\ \GeV,\quad  2 < \eta_{j_1} < 4, \nonumber \\
	&p_{h} > 4\ \GeV,\quad p_{T,h} > 0.25\ \GeV,\quad \Delta R(j_1,h) < 0.5, \nonumber \\
	&p_{T,\ell^\pm} > 20\ \GeV,\quad 2.5 < \eta_{\ell^\pm} < 4.5,\quad
	\Delta R(j_1,\ell) > 0.5,\quad \Delta R(j_1,Z) > \frac{7\pi}{8}\,,
\end{eqnarray}
and jets are reconstructed using the anti-$k_T$ algorithm~\cite{Cacciari:2008gp} with cone size $R = 0.5$.

We consider predictions for $F(z)$ in three different $p_{T,j_1}$ intervals:
$20<p_{T,j_1}<30$\ GeV, $30<p_{T,j_1}<50$\ GeV and $50<p_{T,j1}<100$\ GeV.
We adopt the NNPDF3.1~\cite{NNPDF:2017mvq} PDF set, with $\alpha_s = 0.118$ and
$n_f^\mathrm{max} = 5$: both the PDF and the $\alpha_s$ values are accessed
through the LHAPDF library~\cite{Buckley:2014ana}.
Lastly, the central renormalisation and factorisation scale are chosen to be equal to
\begin{eqnarray}\label{eq:scalemurmuf_lhcb}
	\mu_R=\mu_F=\sqrt{m_Z^2 + \sum_{j\in \text{jets}} p_{T,j}^2}\,,
\end{eqnarray}
where the sum runs over all the jets in the event.
For the fragmentation scale, on the other hand, we choose the central scale
\begin{eqnarray}\label{eq:scalemua_lhcb}
	\mu_D={  R \cdot \ptj}\,,
\end{eqnarray}
where $R=0.5$ is the cone size used in the jet algorithm. This particular choice for the scale $\mu_D$ is motivated by the fact that the hadron in the process definition must be contained within the leading jet, thereby indicating that the resolution scale for the fragmentation process should be related to the jet characteristics. For the numerator, the theoretical uncertainties are determined by the envelope of the common 7-point scale variation, 
by correlating the variation of $\mu_F$ and $\mu_D$ and otherwise independently halving or doubling the scales and omitting the most extreme variations. In the case of the denominator, we determine the $Z+\text{jet}$ cross section at the same central scale as in eq.~\ref{eq:scalemurmuf_lhcb}.


We provide two categories of predictions depending on the hadron types produced inside the jet recoiling against the $Z$-boson.
In section~\ref{subsec:Zh_results_sum} we consider the sum of unidentified charged hadrons, including pion, kaon and proton contributions.
In section~\ref{subsec:Zh_results_pions} we separately consider the pion contribution, the latter being by far the dominant contribution (kaons and protons contribute in a similar way, with the size of their contributions amounting to 20--30\% of the pion contribution).

 
\subsection{Results for unidentified charged hadron production}
\label{subsec:Zh_results_sum}

Predictions for charged hadron production are computed using three different sets of fragmentation functions:
BKK~\cite{Binnewies:1995pt}, NNFF1.1~\cite{Bertone:2018ecm}, DSS07~\cite{deFlorian:2007aj,deFlorian:2007ekg}.
%
BKK and DSS07 have been directly implemented in the NNLOJET infrastructure, 
while NNFF1.1 is available through the LHAPDF interface. From these, only NNFF1.1 provides error sets, thus we refrain from showing the corresponding errors in our results so that the treatment is the same in the three cases, allowing us to compare better their features.
In the following, we provide some details on each of the three fragmentation function sets.

The BKK set~\cite{Binnewies:1995pt} was obtained by fits to LEP and HERA data from the early 90's. The unidentified charged hadron FF is defined as the arithmetic sum of the charged pion and kaon contributions. The parton-to-charged pions and parton-to-charged kaons FFs are provided as simple analytic functions which encode the DGLAP evolution at LO and NLO level.
The NNFF1.1 set~\cite{Bertone:2018ecm} does not consider individual charged hadron contributions, but directly determine an unidentified charged hadron FF through a global NLO fit of a large variety of measurements at $\mathrm{e}^+\mathrm{e}^-$ colliders, the Tevatron and LHC.
In the DSS07 set~\cite{deFlorian:2007aj,deFlorian:2007ekg}, the determination of the unidentified charged hadron FF is related to the FFs for identified pions, kaons and (anti-)protons by imposing the sum rule:
\begin{eqnarray}
\label{eq:FF_sumrule}
D_a^{h^\pm} = D_a^{\pi^\pm} + D_a^{K^\pm} + D_a^{p/\bar{p}} + D_a^{\text{res.}^\pm}
\end{eqnarray}
as a constraint. In particular, $D_a^{\pi^\pm}, D_a^{K^\pm}$ and $D_a^{p/\bar{p}}$ are individually fitted, and then used as input in a subsequent fit to determine the residual charged hadron contribution $D_a^{\text{res.}^\pm}$.
The different FFs are determined through a global NLO fit of data from electron--positron annihilation, proton--proton collisions and deep-inelastic lepton--proton scattering.

\begin{figure}[tp]
	\centering  
	\includegraphics[width=49mm]{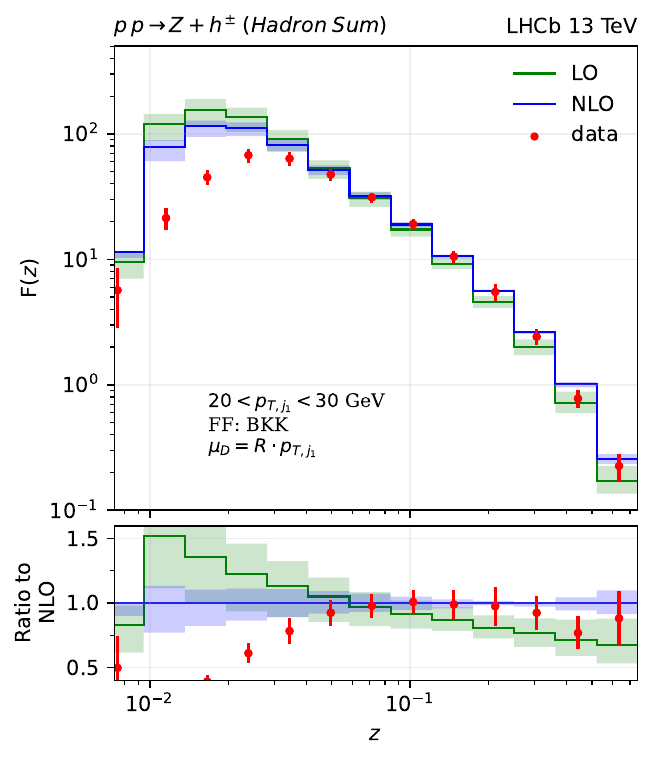}
	\includegraphics[width=49mm]{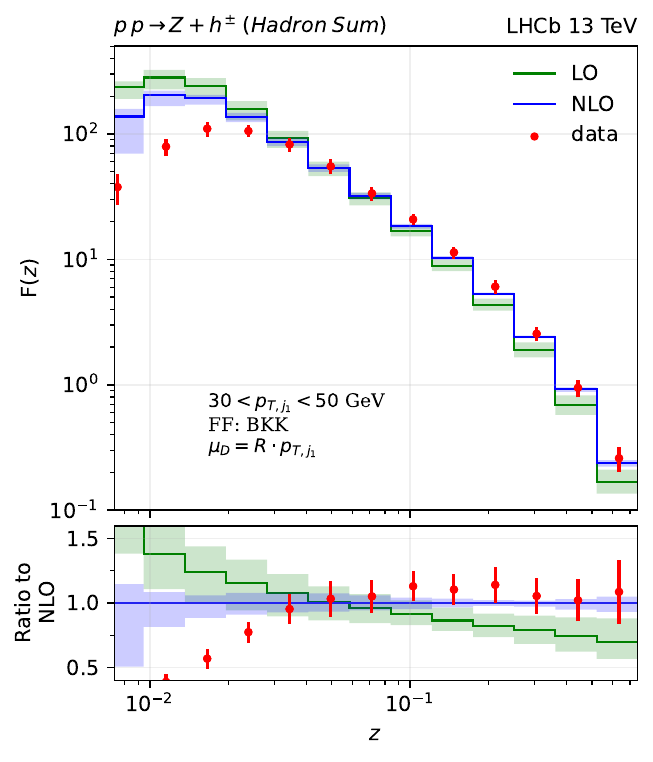}	
	\includegraphics[width=49mm]{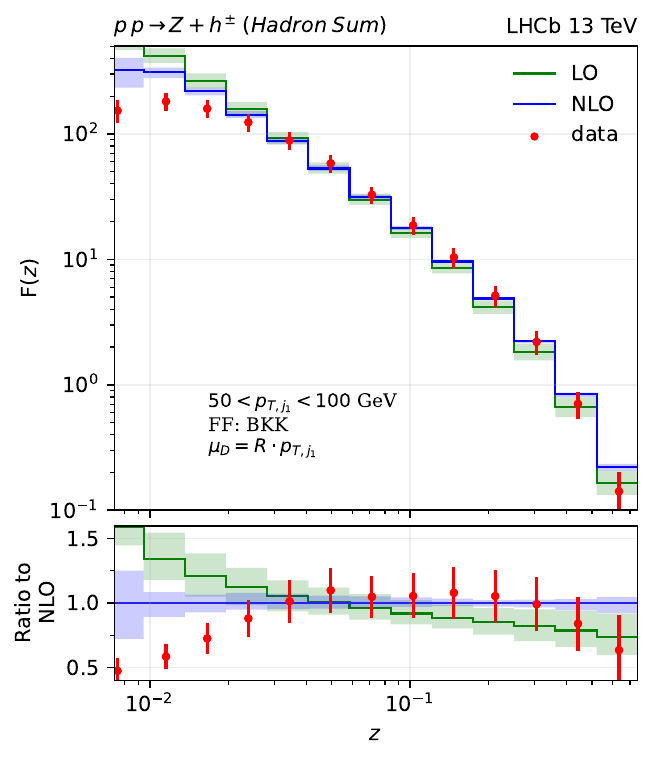}
	\includegraphics[width=49mm]{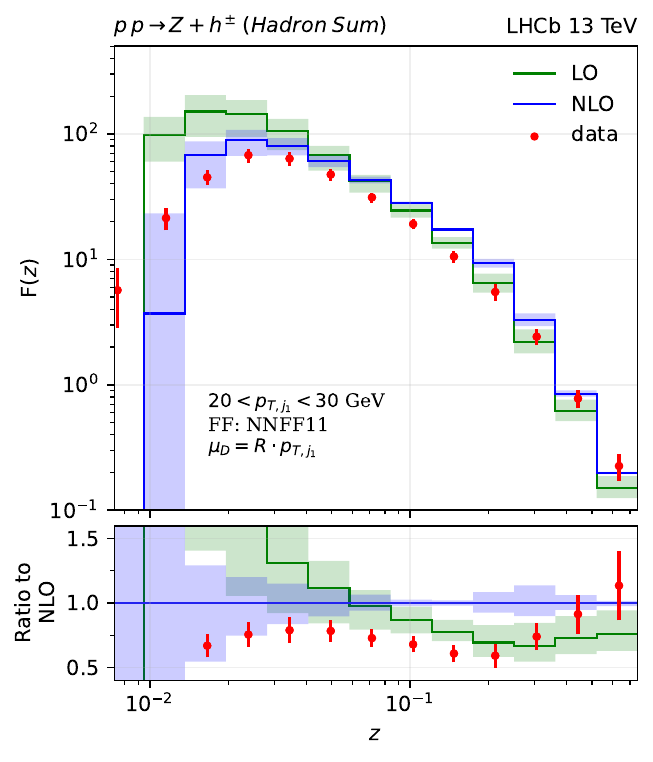}
	\includegraphics[width=49mm]{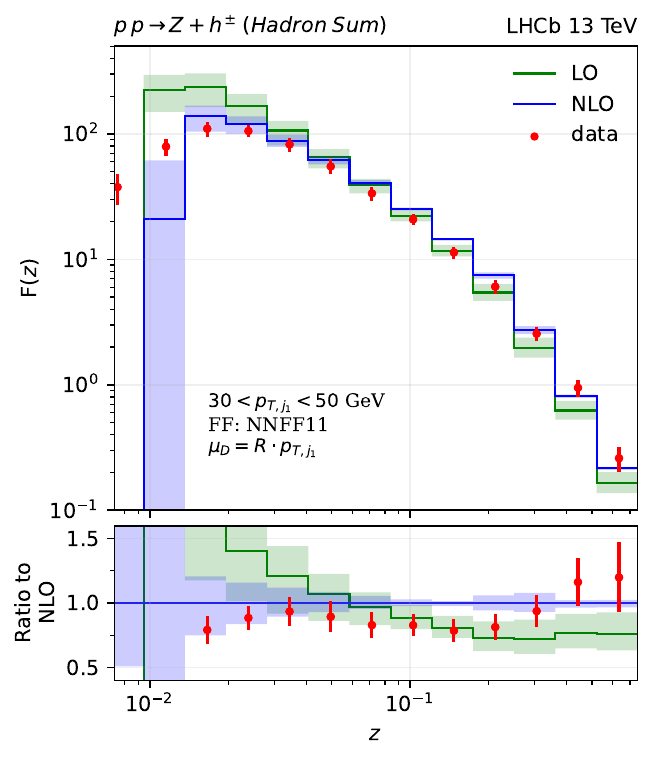}	
	\includegraphics[width=49mm]{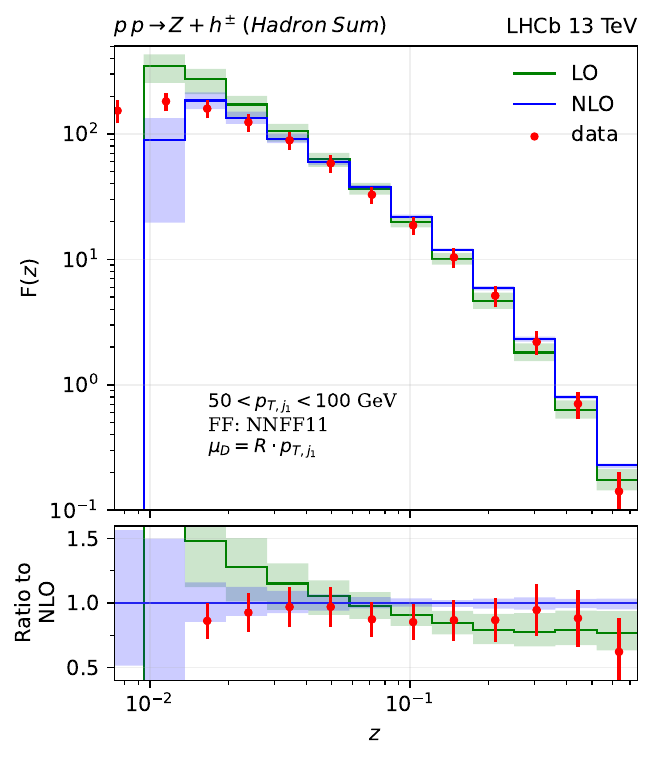}
	\includegraphics[width=49mm]{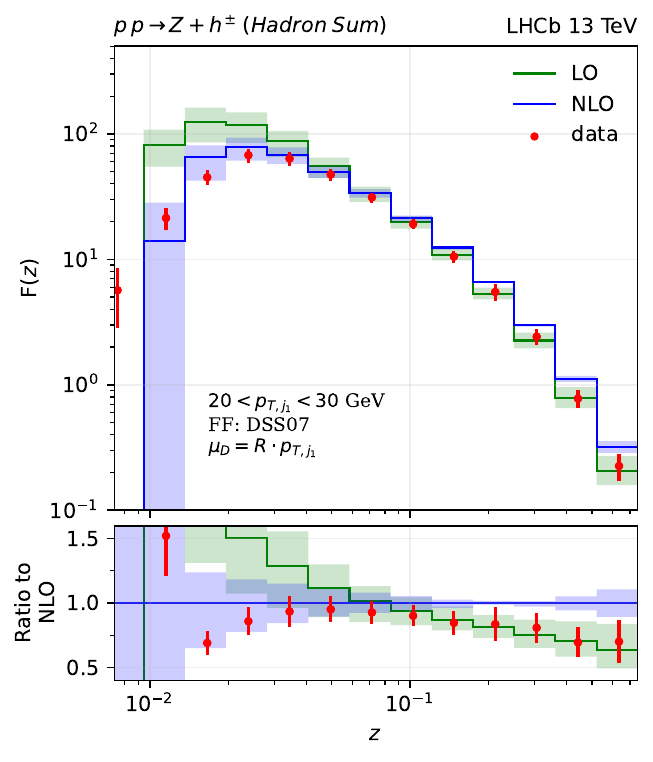}
	\includegraphics[width=49mm]{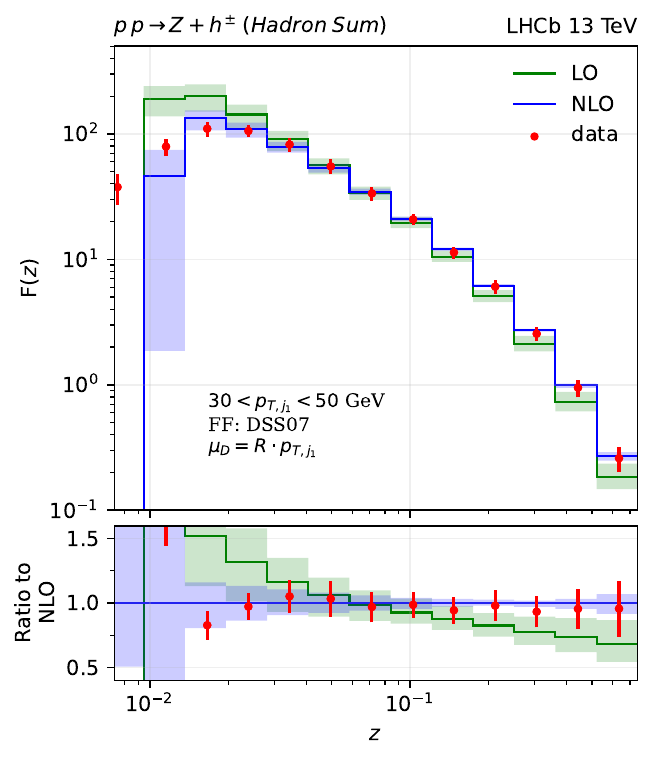}	
	\includegraphics[width=49mm]{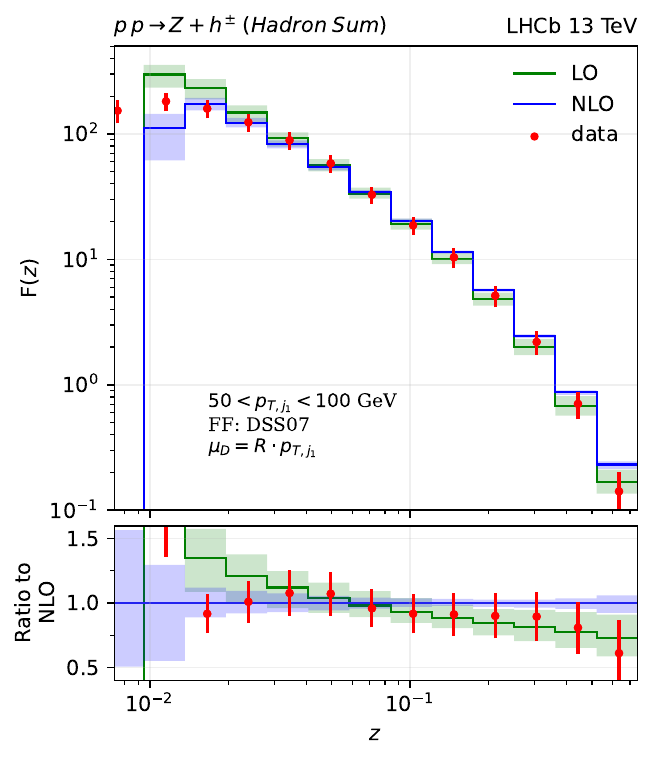}
		
\caption{Comparison of LO (green) and NLO (blue) results for $F(z)$ 
obtained with the BKK (top row), NNFF1.1 (middle row) and DSS07 (bottom row) fragmentation functions with LHCb data (red dots) at 13\ TeV. From left to right, each column corresponds to the $\ptj$ ranges [20, 30], [30, 50] and [50, 100]\ GeV. In each figure, the top panels show the $F(z)$ distributions and the bottom one shows the ratio to NLO. }
\label{fig:z_dscale1_bkk_nnff11}
\end{figure}

Figure~\ref{fig:z_dscale1_bkk_nnff11} presents our predictions for $F(z)$ for three different $p_{T,j1}$ intervals and using the three different FF sets.
Predictions using the BKK parametrization are shown on the top row, NNFF1.1 in the middle row, and DSS07 in the bottom row.
Each row has three sub-figures corresponding to three different $p_{T,j1} $ intervals: [20, 30], [30, 50], [50, 100]\ GeV.
Each sub-figure is made of two panels: the top panel shows the $F(z)$ distributions, and the bottom one shows the ratio to NLO.


For low $z$ values, we do not expect fixed-order perturbative result to be adequate in describing the data,
as this region prone to soft physics, where resummation becomes mandatory.


We first note that when moving from LO to NLO, predictions change in shape, with the size of NLO corrections increasing towards high $z$ values to up to 25\%. We also note a reduction of scale uncertainties.
%
We find that predictions using the BKK and DSS07 charged hadron FF best describe the data: for mid $z$ values, both sets show alignment between NLO and data for the three $\ptj$ bins, with the exception of the [30, 50]\ GeV $p_{T,j_1}$ interval where BKK underestimates the data; for high $z$ values, NLO predictions with both BKK and DSS07 fail to describe the data in the highest $p_{T,j_1}$ interval, with the LO predictions being closer to the experimental data points. In the lower $p_{T,j_1}$ cases, the agreement at high $z$ values improves, most notably in the middle [30, 50]\ GeV $p_{T,j_1}$ interval.
In contrast, the results computed using NNFF1.1 are consistently above the data. 
Furthermore, the shape of the distributions differ for each FF.



\begin{figure}[tp]
	\centering  
	\includegraphics[scale=0.45]{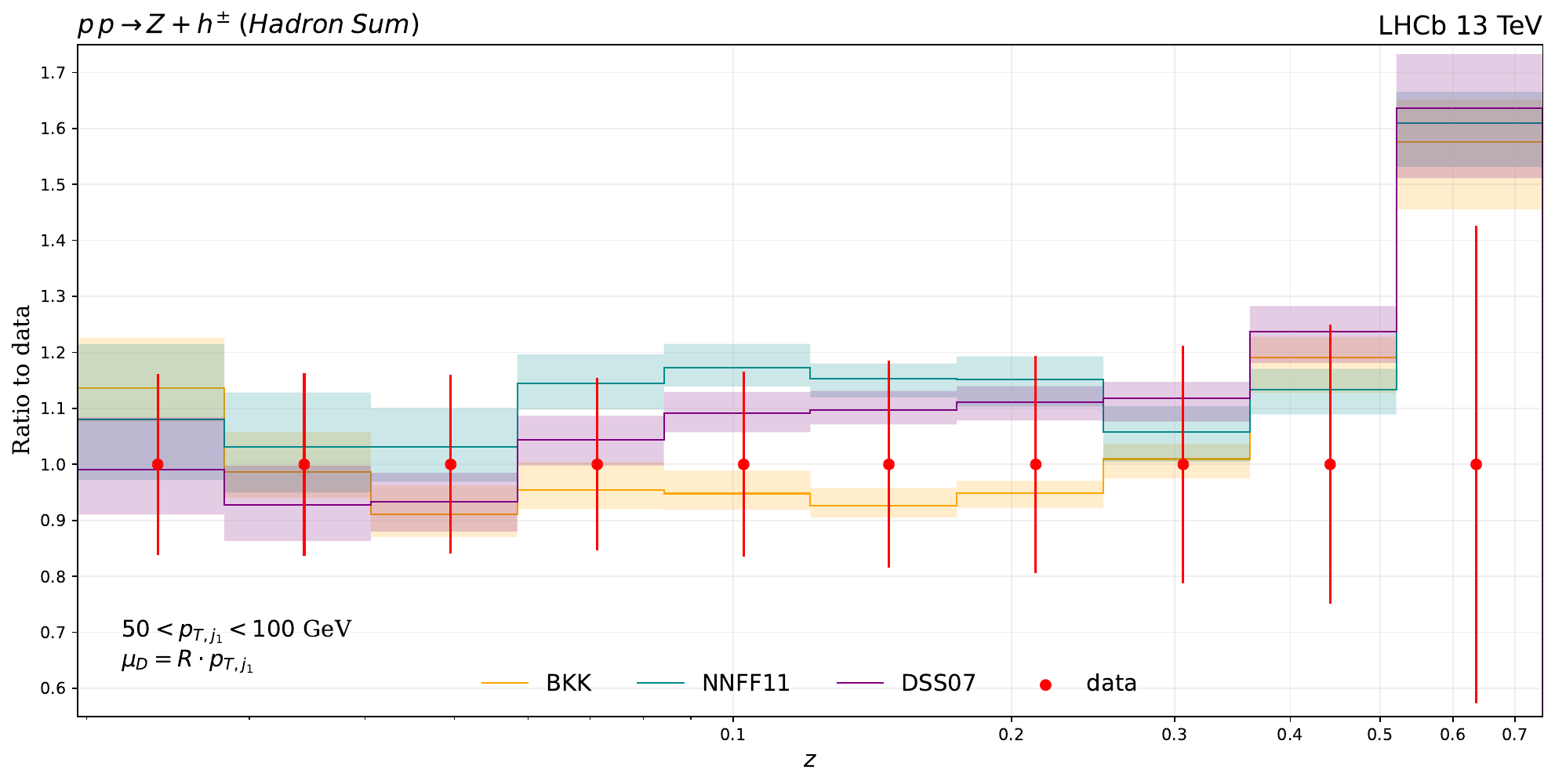}
	\caption{Comparison of the NLO results obtained for the different sets of FFs shown in Fig.~\ref{fig:z_dscale1_bkk_nnff11} for $\ptj\in$ [50, 100]\ GeV. Note that when comparing with Fig.~\ref{fig:z_dscale1_bkk_nnff11}, the lower $z$ edge is here increased to $z\sim 0.02$.
	}
	\label{fig:z_dscale1_zoom_zbin3}
\end{figure}

It is interesting to have a closer look into the hardest available region, which corresponds to the third interval in transverse momentum, i.e.\ we focus on the $50 < \ptj < 100$\ GeV bin and restricting the $z$ range to values $z>0.02$. This is shown in Fig.~\ref{fig:z_dscale1_zoom_zbin3}, where we plot the ratio to data of all three NLO predictions for $F(z)$.
We observe a compatibility of the results for the three FFs in the highest and lowest bins of Fig.~\ref{fig:z_dscale1_zoom_zbin3}  within the shown range:
for $z < 0.06$, data and theory are compatible within uncertainties, while for $z > 0.35$ theory predictions start to be
above the data, with larger experimental uncertainty.
In the intermediate range of z, BKK and DSS07 seem to provide a better description of data compared to NNFF1.1.

\subsection{Results for charged pion production}
\label{subsec:Zh_results_pions}

We now move on to $Z$-tagged events associated with the production of charged pions inside jets.
We adopt the pion FFs from BKK~\cite{Binnewies:1995pt}, NNFF1.0~\cite{Bertone:2017tyb} and DSS07~\cite{deFlorian:2007aj}.
Note that in this context, we adopt the NNFF1.0 set instead of NNFF1.1 above: this is due to the fact that NNFF1.1 only provides an unidentified charged hadron set, whereas NNFF1.0 provides individual pions, kaons and (anti-)proton FF sets.
As for NNFF1.1, also NNFF1.0 is provided via the LHAPDF interface.

\begin{figure}[tp]
	\centering  
	\includegraphics[width=49mm]{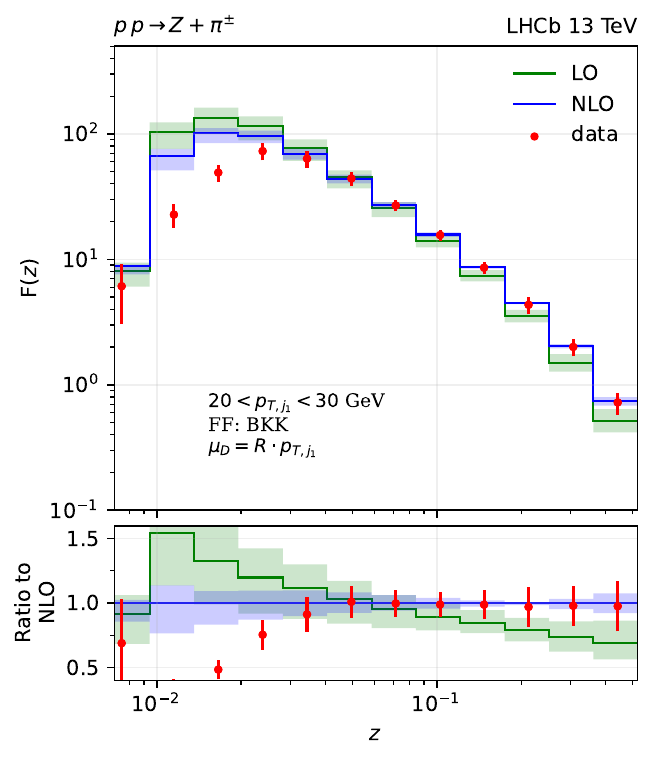}
	\includegraphics[width=49mm]{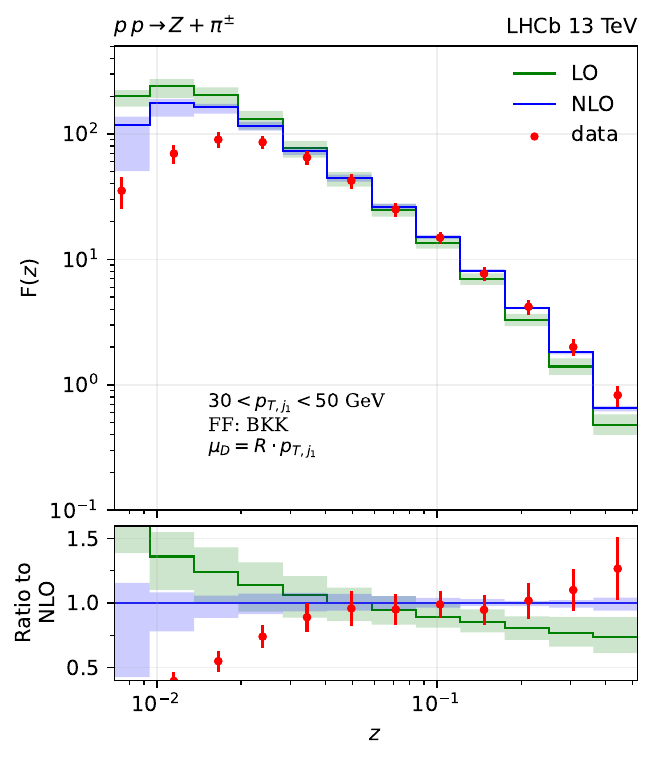}	
	\includegraphics[width=49mm]{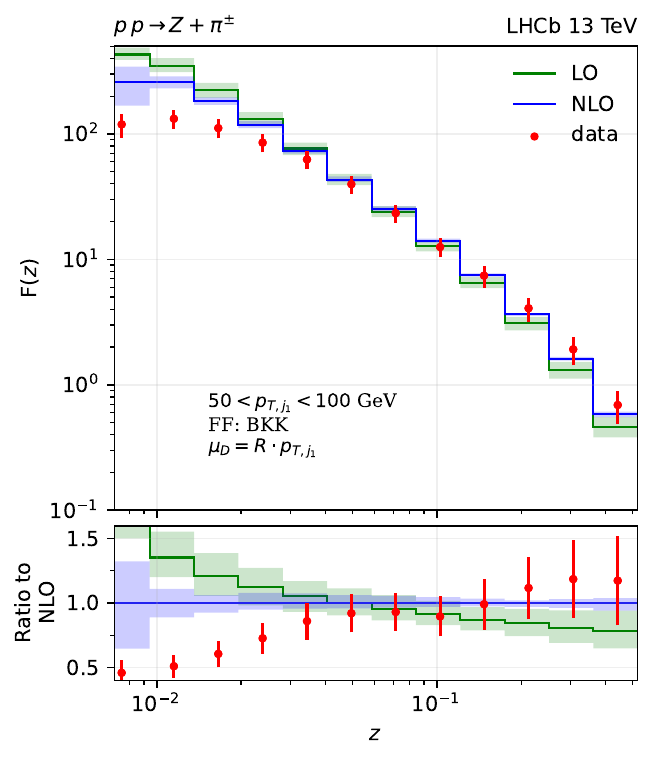}
	\includegraphics[width=49mm]{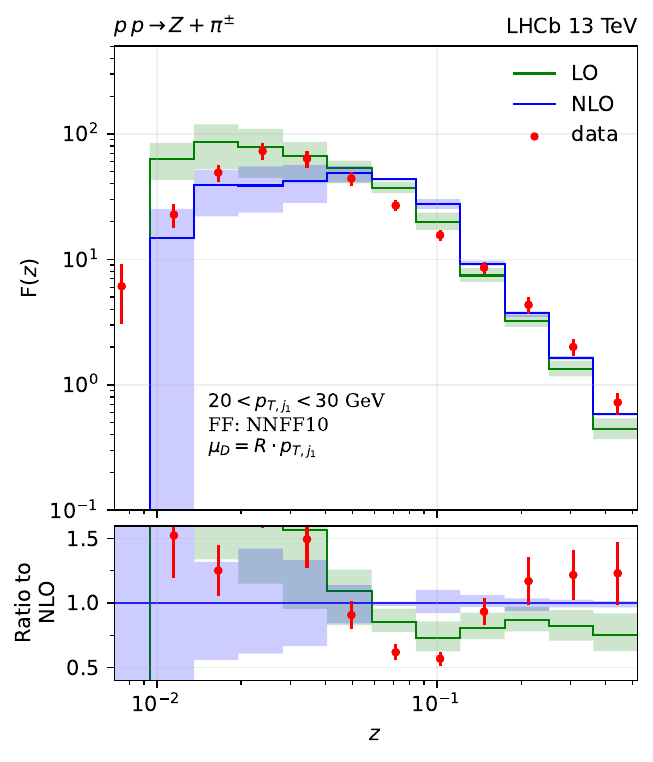}
	\includegraphics[width=49mm]{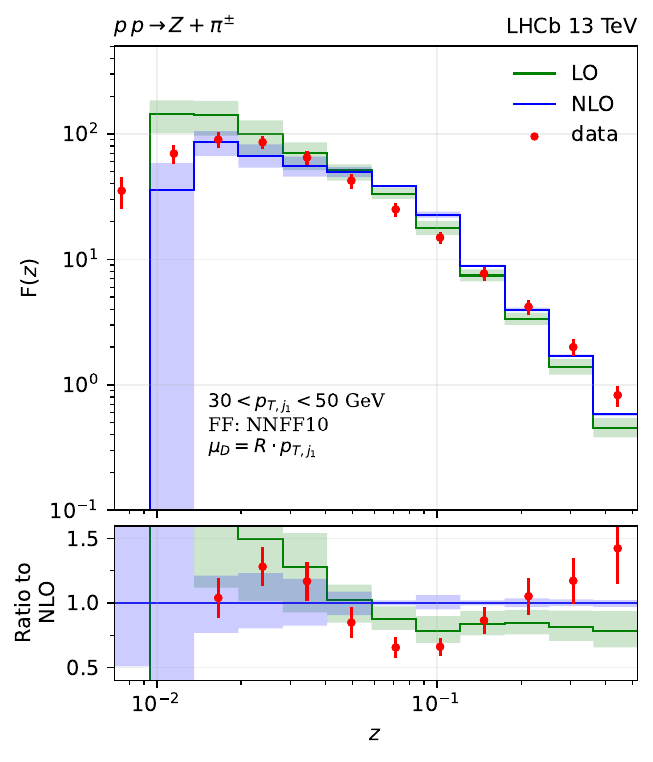}	
	\includegraphics[width=49mm]{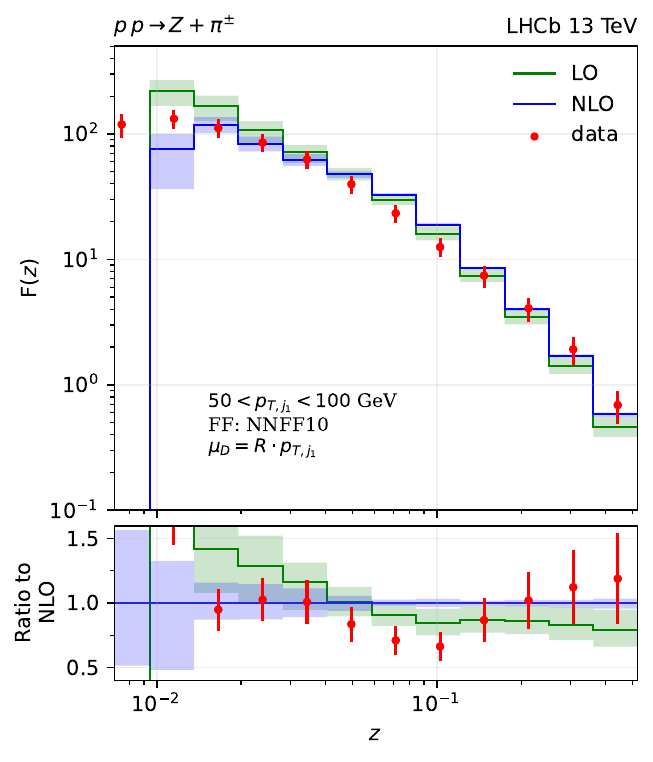}
	\includegraphics[width=49mm]{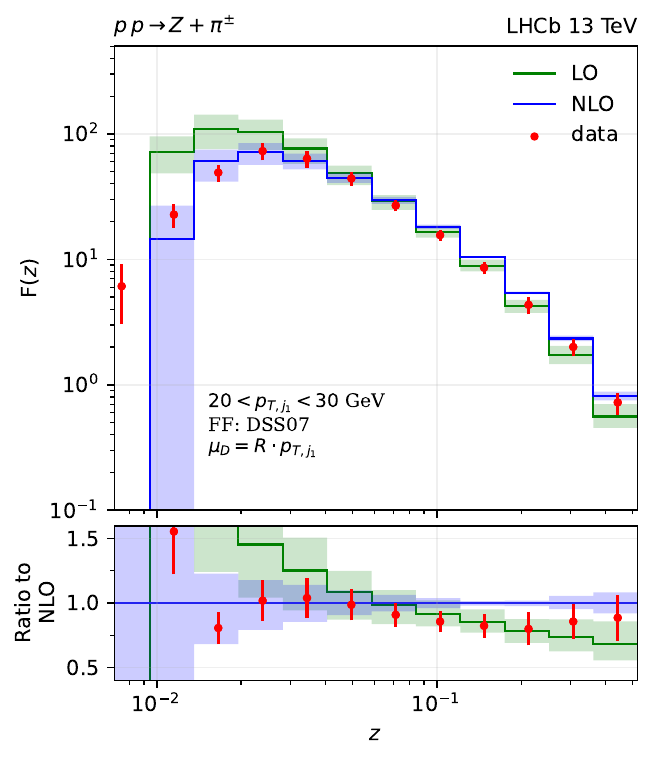}
	\includegraphics[width=49mm]{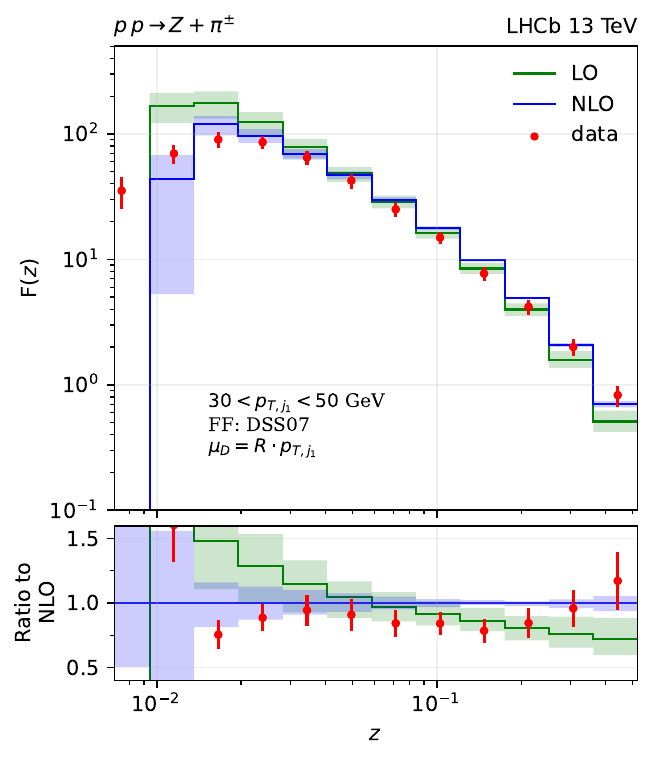}	
	\includegraphics[width=49mm]{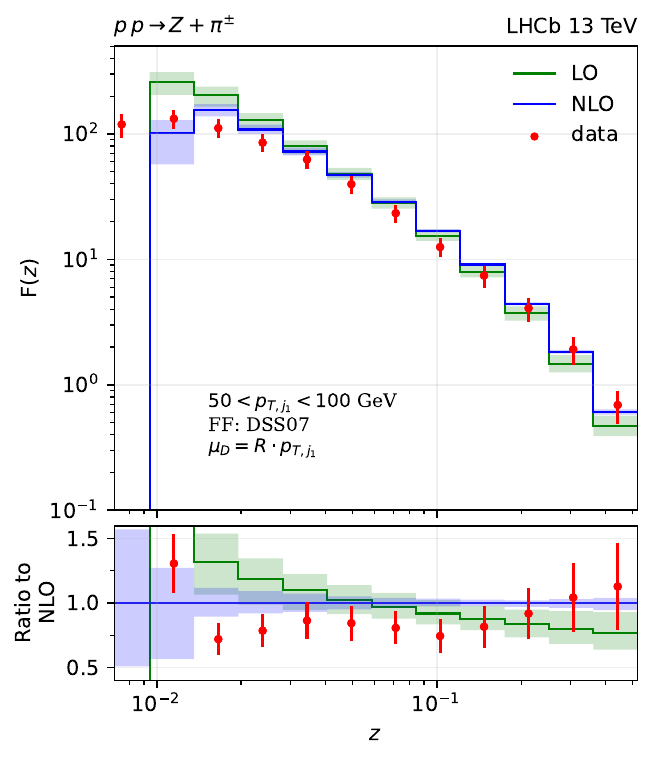}
\caption{Comparison of LO (green) and NLO (blue) results for $F(z)$ obtained with the BKK (top row), NNFF1.0 (middle row) and  DSS07 (bottom row) fragmentation functions for pions with LHCb data (red dots) at 13\ TeV. From left to right, each column corresponds to the $\ptj$ ranges [20, 30], [30, 50] and [50, 100]\ GeV. The top panel shows the $F(z)$ distributions and the bottom one shows the ratio to NLO.}
\label{fig:z_dscale1_bkk_nnff_dss_sephad_pions}
\end{figure}

Our results for $F(z)$ using the BKK, NNFF1.0 and DSS07 pion FFs are shown in Fig.~\ref{fig:z_dscale1_bkk_nnff_dss_sephad_pions} using the same layout as in Fig.~\ref{fig:z_dscale1_bkk_nnff11}.

%
When comparing Fig.~\ref{fig:z_dscale1_bkk_nnff_dss_sephad_pions} to the analogous plot for unidentified hadrons in Fig.~\ref{fig:z_dscale1_bkk_nnff11}, we notice similar features.
However, compared to Fig.~\ref{fig:z_dscale1_bkk_nnff11}, in the mid $z$ region, the results computed using NNFF1.0 and DSS07 pion FFs show a tendency to overshoot the data
, whereas the results using the BKK FF describe the data in a satisfactory manner.  
In the high $z$ region, BKK and DSS07 seem to better describe the data compared to the unidentified hadron case, with the BKK set offering the best description.
In all three $p_{T,j_1}$ intervals, for intermediate to high values of $z$, we find that the NLO corrections give positive corrections of up to 25\% with respect to the LO results.



\begin{figure}[tp]
	\centering  
	\includegraphics[scale=0.45]{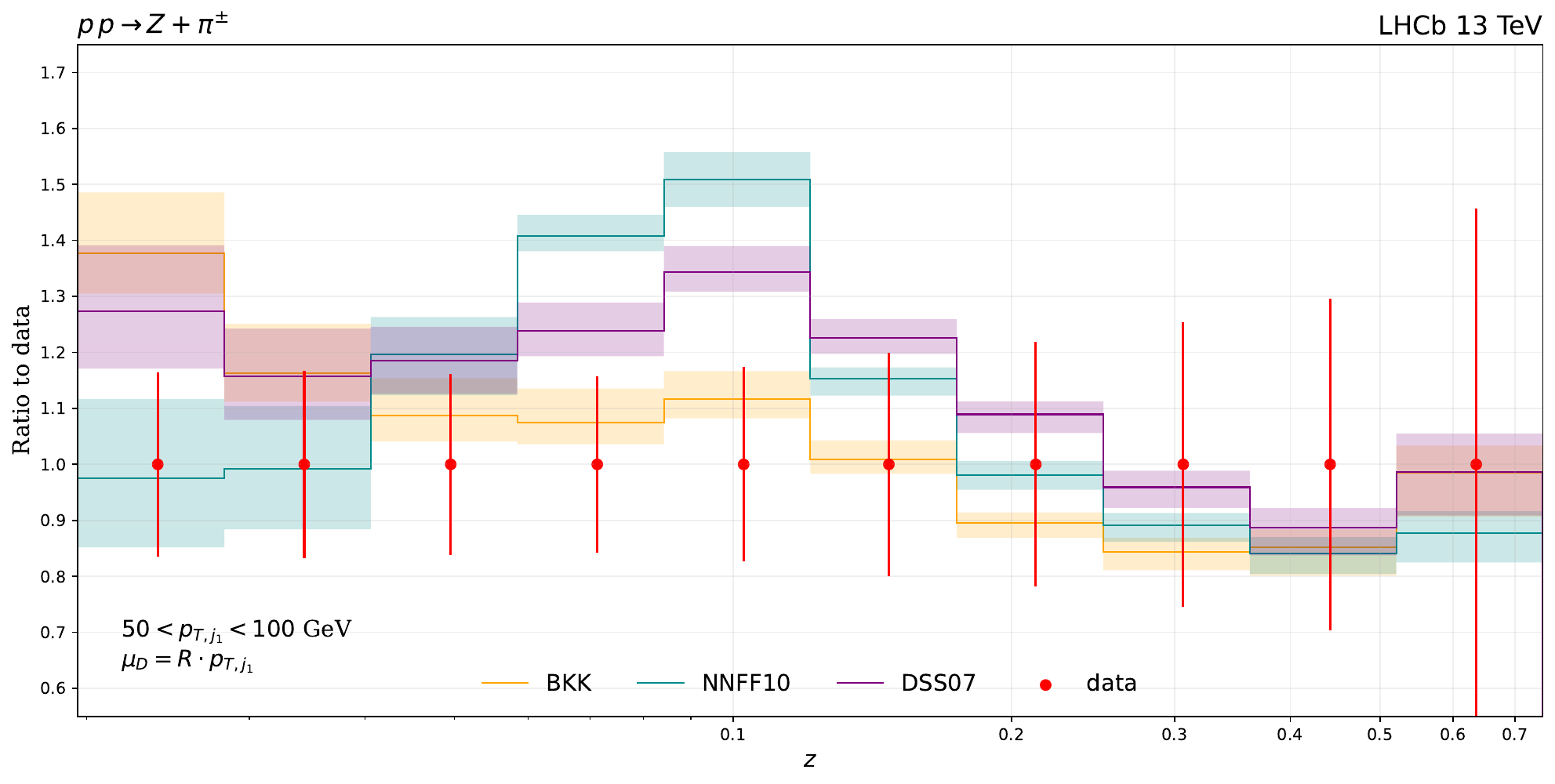}
	\caption{Comparison of the NLO results obtained for the different sets of fragmentation functions shown in Fig.~\ref{fig:z_dscale1_bkk_nnff_dss_sephad_pions} in the region $\ptj\in [50, 100]$\ GeV. Note that when comparing with Fig.~\ref{fig:z_dscale1_bkk_nnff_dss_sephad_pions}, the lower $z$ edge is here increased to $z\sim 0.02$.}
	\label{fig:z_dscale1_pions_zbin3}
\end{figure}

In order to further study the impact of the different FFs on the theory predictions, in Fig.~\ref{fig:z_dscale1_pions_zbin3} we compare the $F(z)$ distributions for pion production with the three FF sets in the highest $\ptj$ bin.
Predictions using BKK fragmentation functions provide the best description of data, while the three sets are compatible among each other in the high $z$-region.


In summary, from our analysis at NLO level, we find that the three FFs considered (BKK, NNFF1.0 or NNFF1.1, DSS07) show qualitative differences in the description of LHCb hadron-in-jet data, with none of the three sets able to describe all the kinematical bins equally well.
It is likely that this dataset is able to offer important constraints on FFs when included in global fits.
Indeed, such LHCb dataset has been included in a very recent NLO global analysis of FFs~\cite{Gao:2024nkz}: it has been shown how these hadron-in-jet data are useful to better determine the gluon-to-hadron fragmentation functions, which other datasets leave largely unconstrained.
As our goal here is to study how publicly available FFs obtained through fits of one-particle inclusive datasets are able to describe this new exclusive dataset, we refrain from presenting results obtained using the FFs of~\cite{Gao:2024nkz}.


\section{\texorpdfstring{$W$}{W} boson in association with a charmed hadron}
\label{sec:W+D}
In this section, we present predictions for the associated production of a $W$ boson with a  $D^{(*)}$ meson. More specifically, the following processes are considered: $W^+ + D^{-}$, $W^- + D^{+}$, $W^+ + D^{*-}$ and $W^- + D^{*+}$, with a leptonically decaying $W$ boson. The predictions are compared to data from the ATLAS experiment at 13\ TeV~\cite{ATLAS:2023ibp}.
The cross sections measured in~\cite{ATLAS:2023ibp} are differential either in the transverse momentum $p_{T,h}$ of the $D^{(*)}$ hadron or in the absolute value of the pseudo-rapidity $|\eta_\ell|$ of the lepton from the $W$-boson decay.

In the following, we first provide details about the kinematical cuts and the numerical setup adopted, and we discuss our choice of $D^{(*)}$-meson fragmentation functions. We then divide the presentation in the rest of the section according to the observable considered.
\subsection{Fiducial cuts and numerical setup}
\label{subsec:observable_atlas}
The fiducial region of the ATLAS measurement is defined as follows:
\begin{equation}\label{eq:fiducialcuts_atlas13}
  p_{T,h} > 8\ \GeV,\quad |\eta_{h}| < 2.2,\quad p_{T,\ell} > 30\ \GeV,\quad |\eta_{\ell}| < 2.5,
\end{equation}
without any requirement on the presence of reconstructed jets.
As in section~\ref{sec:Z+h}, we provide predictions using the NNPDF3.1 PDF set~\cite{NNPDF:2017mvq} with values of $\alpha_s = 0.118$ and $n_f^\mathrm{max}=5$.
The central renormalisation, factorisation and fragmentation scales are chosen to be equal to the transverse mass of the $W$ boson:
\begin{eqnarray}\label{eq:scale_atlas}
	\mu_R=\mu_F=\mu_D=m_T^W\,.
\end{eqnarray}
We estimate theoretical uncertainties with a 7-point scale variation using the same scheme as for the $Z$+hadron results in section ~\ref{sec:Z+h}, i.e.\ by maintaining $\mu_F=\mu_D$ while varying $\mu_R$ by separately halving or doubling their values, and by discarding pairs of extreme variations. The electroweak parameters are computed in the $G_\mu$-scheme with values
 \begin{eqnarray}\label{eq:ew_param}
 	&M_Z=91.1876\ \GeV,\quad \Gamma_Z = 2.4952\ \GeV\,,
  \nonumber \\ 
 	&M_W=80.379\ \GeV,\quad \Gamma_W=2.085\ \GeV, \quad	G_\mu = 1.1663787\cdot 10^{-5}\ \GeV^{-2}\,.
 \end{eqnarray}
Furthermore, we use a non-diagonal CKM matrix with Wolfenstein parametrisation given by $\lambda =0.2265$, $A=0.79$, $\bar \rho=0.141$ and $\bar \eta =0.357$~\cite{ParticleDataGroup:2020ssz}.
 
In our predictions, we compare the results obtained using two $D^{(*)}$-meson fragmentation function sets, CNO~\cite{Cacciari:2005uk} and KKKS08~\cite{Kneesch:2007ey}.
We first provide some details concerning the individual sets, given that they feature quite substantial differences in the methodology adopted for their determination.

The CNO set is derived by exploiting the perturbative fragmentation function formalism for heavy quarks~\cite{Mele:1990yq}, in order to resum large logarithms of a hard scale over the charm quark mass.
Namely, the NLO perturbative initial condition at the charm-mass scale---further supplemented with next-to-leading-logarithmic soft gluon resummation---is evolved with a NLO DGLAP evolution to the higher scales.
A three-parameter non-perturbative component (Colangelo--Nason form supplemented with a hard term) is added on top of the perturbative result, with values for the parameters obtained through a fit to CLEO and BELLE data. See~\cite{Cacciari:2005uk} for additional details.

Instead, the KKKS08 set is obtained through a global fit of BELLE, CLEO, ALEPH and OPAL data in a way similar to light-hadron FFs fits, without any perturbative input.
Namely, both the charm and bottom FFs are parametrised with a Bowler-like form with three parameters at their respective mass scales, and then they are evolved to higher scales with a NLO DGLAP evolution.
Two variants of the KKKS08 are presented in~\cite{Kneesch:2007ey}: a zero-mass (ZM) variant, where all quark masses are neglected, and a general-mass (GM) approach, which includes $b$- and $c$-quark finite-mass corrections.
In the following, we will present predictions for the GM variant only, in order to have two $D^{(*)}$-meson FF sets as different as possible.


Lastly, in our predictions we must take into account the {\em opposite sign minus same sign} (OS$-$SS) prescription applied in the measurement.
Such a prescription consists in subtracting the contributions where the $W$ boson and the $D$ hadron have the same sign (SS) to the contributions where they have opposite sign (OS).
In order to consistently apply such a prescription in our theoretical predictions, we first compute observables for all the sign combinations of $W$ and $D^{(*)}$ and then we perform the OS$-$SS subtraction on the resulting distributions and fiducial cross sections. For example
\begin{eqnarray}\label{eq:osss_def_example}
  \sigma _{W^-+D^+}^{\text{OS}-\text{SS}} \equiv \sigma _{W^-+D^+}-\sigma _{W^-+D^-}
\end{eqnarray}
is the OS$-$SS fiducial cross section for $W^- +D^+$ production.
Note that due the fact that the charm-to-$D^-$ fragmentation function is small but non-zero, the SS piece already contributes at LO. This is in contrast to predictions for $W+c\text{-jet}$ production~\cite{Gehrmann-DeRidder:2023gdl}, where SS contributions only contribute starting from NLO.

\subsection{Results for the \texorpdfstring{$|\eta_\ell|$}{|ηl|} observable}
\label{subsubsec:WD_yl}

\begin{figure}[tp]
	\centering  
	\includegraphics[width=65mm]{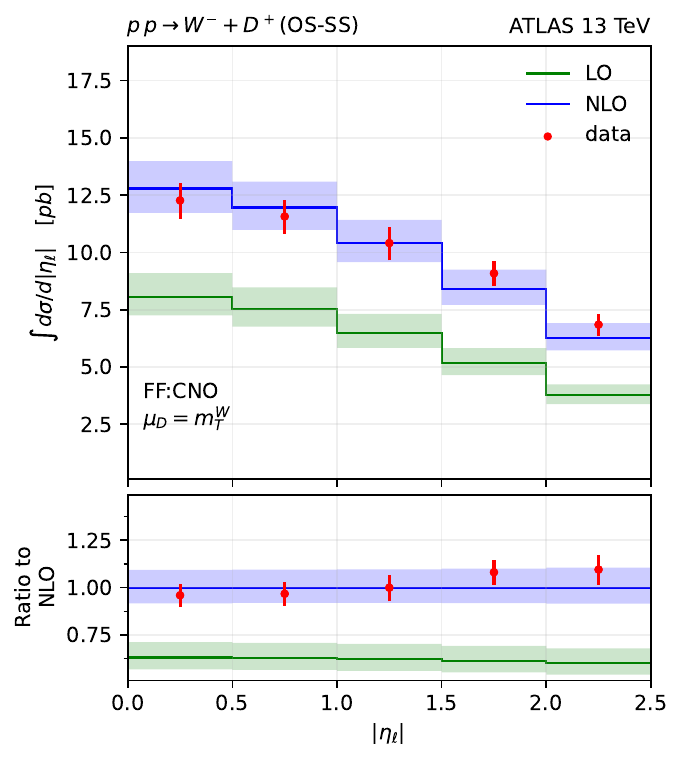}
	\includegraphics[width=65mm]{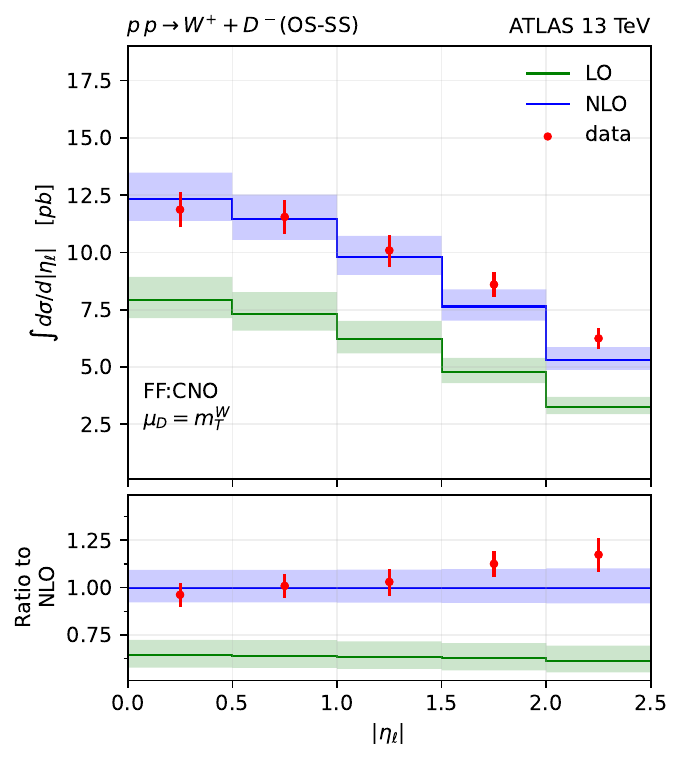}	
	\includegraphics[width=65mm]{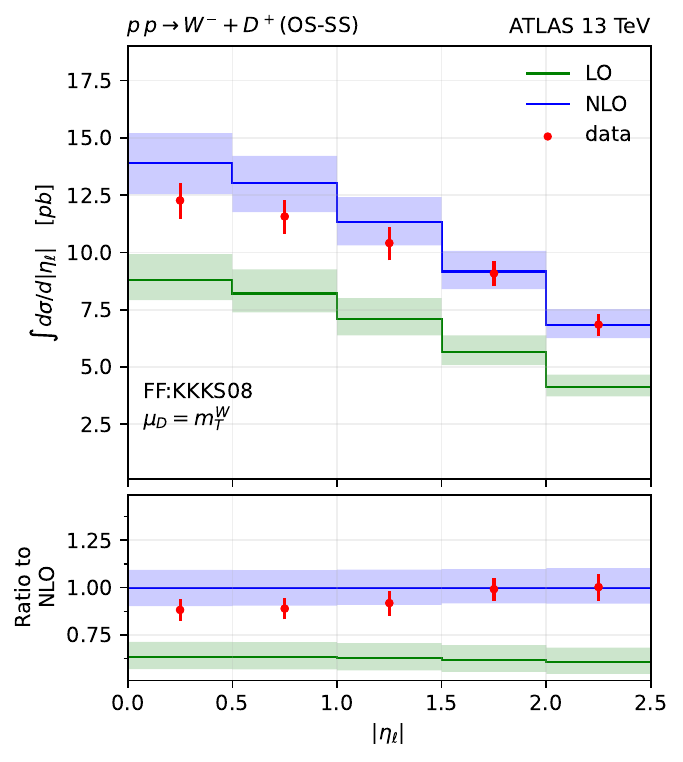}
	\includegraphics[width=65mm]{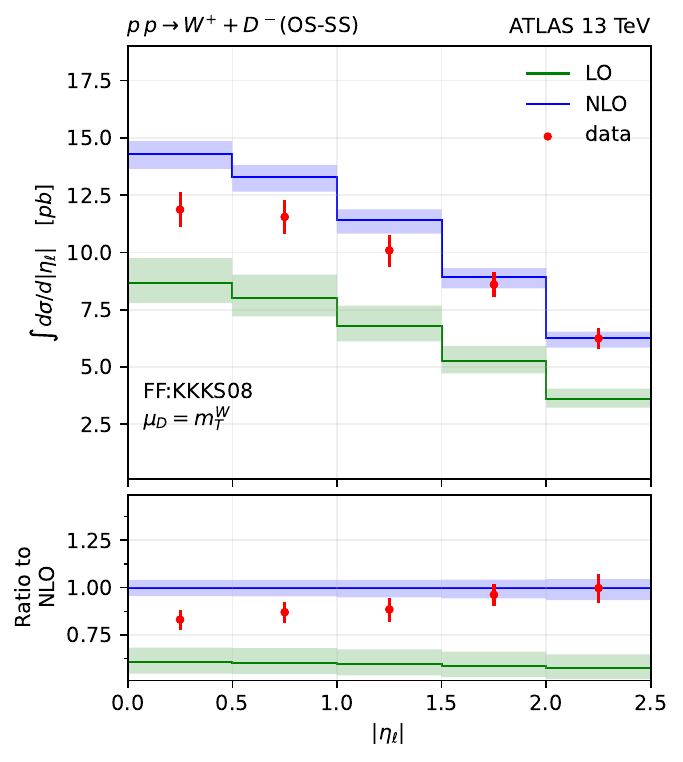}
	\caption{Comparison of LO (green) and NLO (blue) predictions with data (red). The $|\eta_{\ell}|$ distribution is integrated over each differential bin and is shown for all the $D$-hadron cases. The top plots have been produced using the CNO fragmentation function, while the bottom plots using  KKKS08. The left and right columns show the results for $W^-$ and $W^+$, respectively. All include the OS$-$SS prescription.}
	\label{fig:wd_yl}
\end{figure}

\begin{figure}[tp]
	\centering  
	\includegraphics[width=65mm]{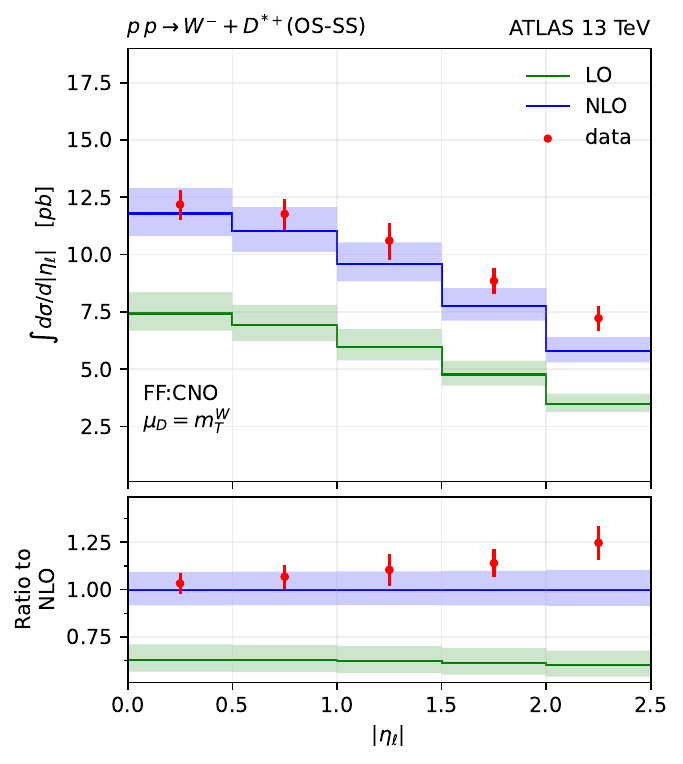}
	\includegraphics[width=65mm]{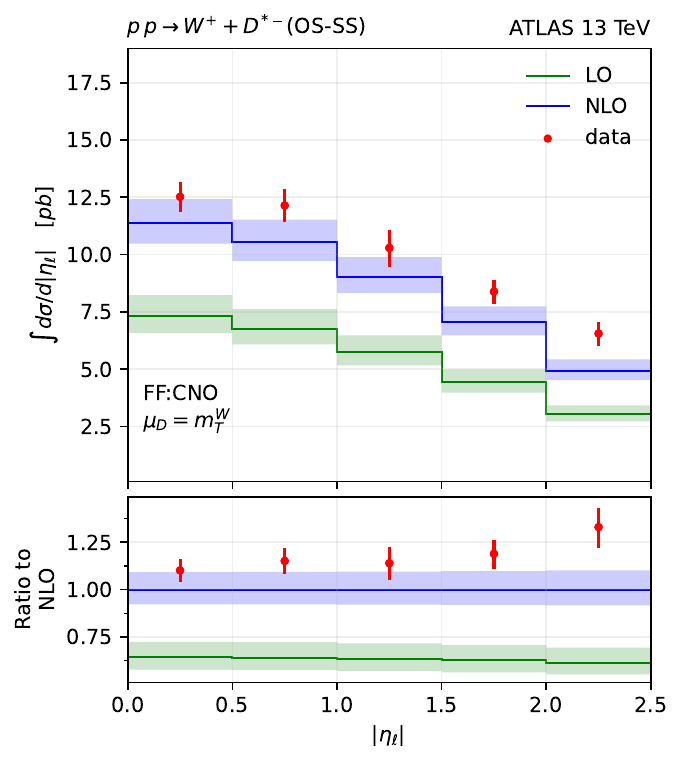}	
	\includegraphics[width=65mm]{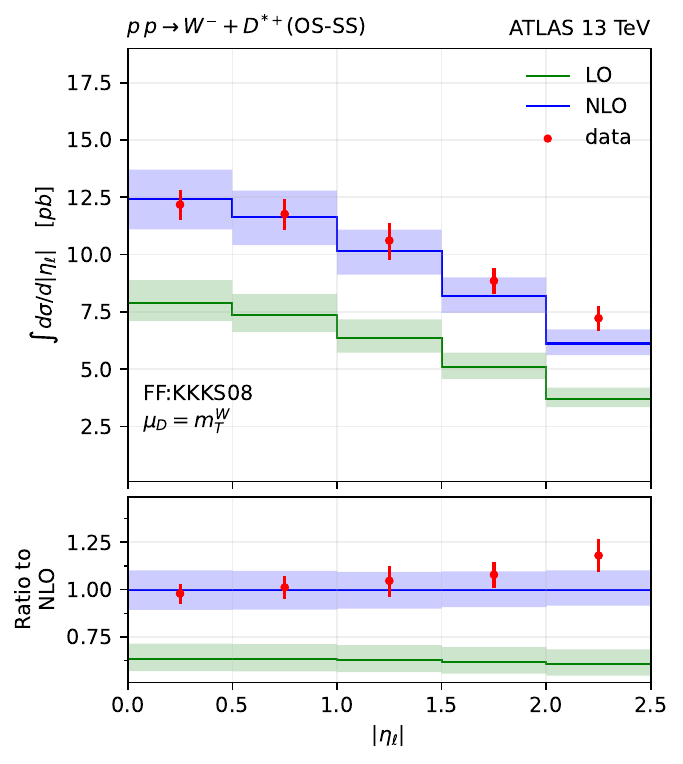}
	\includegraphics[width=65mm]{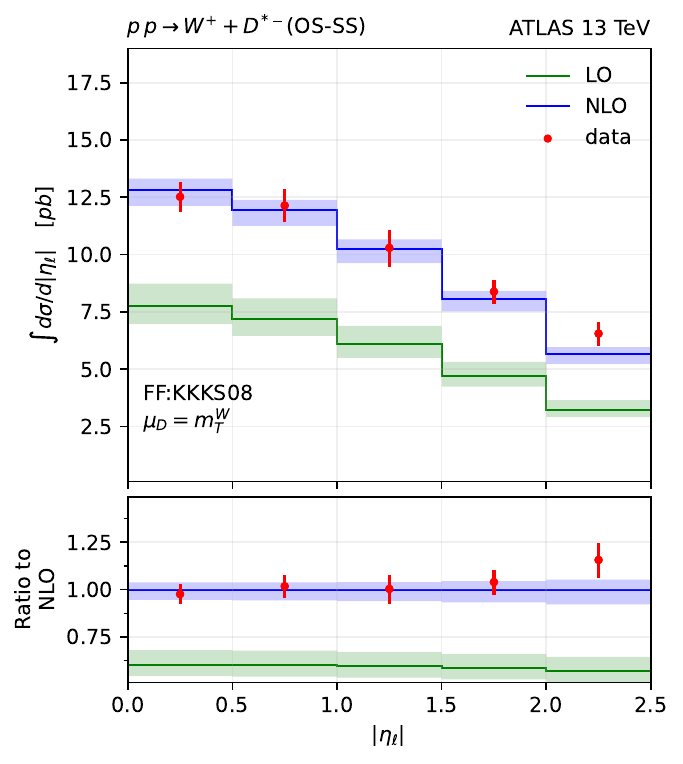}
	\caption{Comparison of LO (green) and NLO (blue) predictions with data (red). The $|\eta_{\ell}|$ distribution is integrated over each differential bin and is shown for all the $D^*$-hadron cases. The top plots have been produced using the CNO fragmentation function, while the bottom plots using  KKKS08. The left and right columns show the results for $W^-$ and $W^+$, respectively. All include the OS$-$SS prescription.}
	\label{fig:wdstar_yl}
\end{figure}

We first consider predictions for differential distributions in $|\eta_\ell|$.
Results for $W$-boson plus $D$-hadron and $W$-boson plus $D^{*}$-hadron production are shown in Fig.~\ref{fig:wd_yl} and Fig.~\ref{fig:wdstar_yl}, respectively.
In each figure, the OS$-$SS results for $W^-+D^+$ ($W^-+D^{*+}$) and $W^++D^-$ ($W^++D^{*-}$) are given on the left and on the right column respectively.
The top row contains plots with predictions using the CNO set, whereas the bottom row features plots adopting the KKKS08 set.
Each plot is divided into two panels, showing the distributions in absolute value (top panel) and their ratio to NLO (bottom panel).

%
We first note that the NLO correction is about $40\%$ and quite flat across the entire $|\eta_\ell|$ spectrum. The results obtained with the CNO fragmentation functions describe the data better in the central region, while the distributions obtained with KKKS08 instead show better agreement in the forward region.
On the other hand, in the $D^*$-hadron case presented in Fig.~\ref{fig:wdstar_yl}, we observe that the  $|\eta_\ell|$ distributions are fairly similar for the two sets of FF. Also in Fig.~\ref{fig:wdstar_yl} NLO corrections are very flat in $|\eta_\ell|$ and amount to about $40\%$. However, the $W^-+D^{*+}$ prediction is in better agreement with the data with respect to the $W^++D^{*-}$ process, especially for the CNO fragmentation function. Finally, the last rapidity bin in the forward region is poorly described in both processes by both FFs.
Overall, we can see a clear improvement in the quality of description of data moving from LO to NLO.

\begin{figure}[tp]
	\centering  
	\includegraphics[scale=0.45]{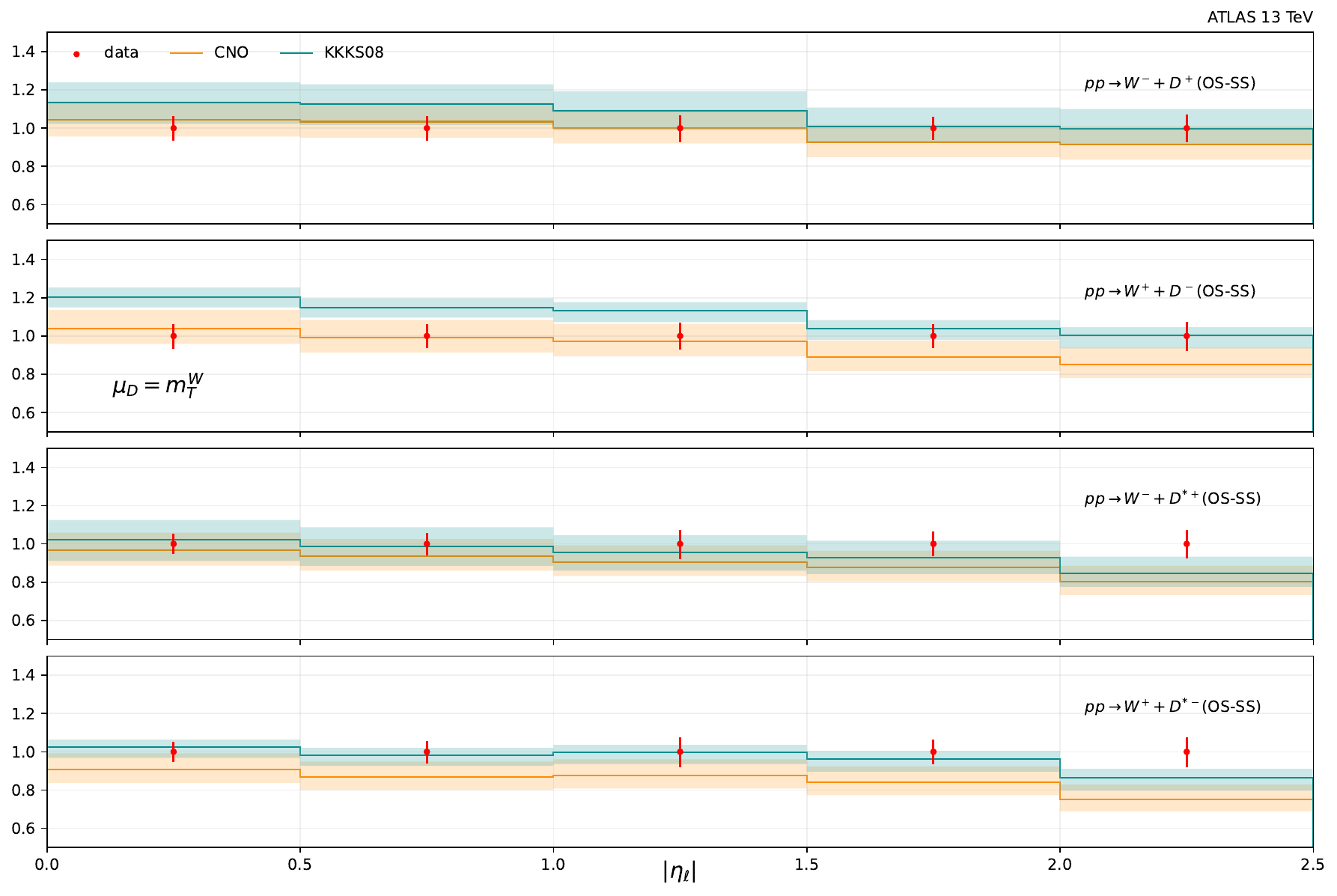}	
	\caption{Comparison of the NLO results obtained for the different sets of fragmentation functions shown in Fig.~\ref{fig:wd_yl} and Fig.~\ref{fig:wdstar_yl}.
          The plots correspond to the ratio to data of the $|\eta_\ell|$ distribution for $D$- and $D^*$-hadrons.}
	\label{fig:squeezed_yl}
\end{figure}

A direct comparison between the predictions obtained with the CNO and the KKKS08 fragmentation functions is given in Fig.~\ref{fig:squeezed_yl} for the $|\eta_\ell|$ distribution.
The plot is composed by four panels, each one showing the ratio of the corresponding NLO distribution to data, for a different process.
From Fig.~\ref{fig:squeezed_yl} we can better assess the impact of changing the fragmentation function between CNO and KKKS08 on the $|\eta_\ell|$ distribution which is fairly small, as expected.
As we already noticed from the $D$-hadron plots in Fig.~\ref{fig:wd_yl}, the KKKS08 distribution slightly overshoots the data in the central region.
We also notice that both the distributions are below the data in the forward region for the $D^*$-hadron case, but KKKS08 is closer to data.

\subsection{Results for the \texorpdfstring{$p_{T,h}$}{pTh} observable}
\label{subsubsec:WD_pthad}
\begin{figure}[tp]
	\centering  
	\includegraphics[width=65mm]{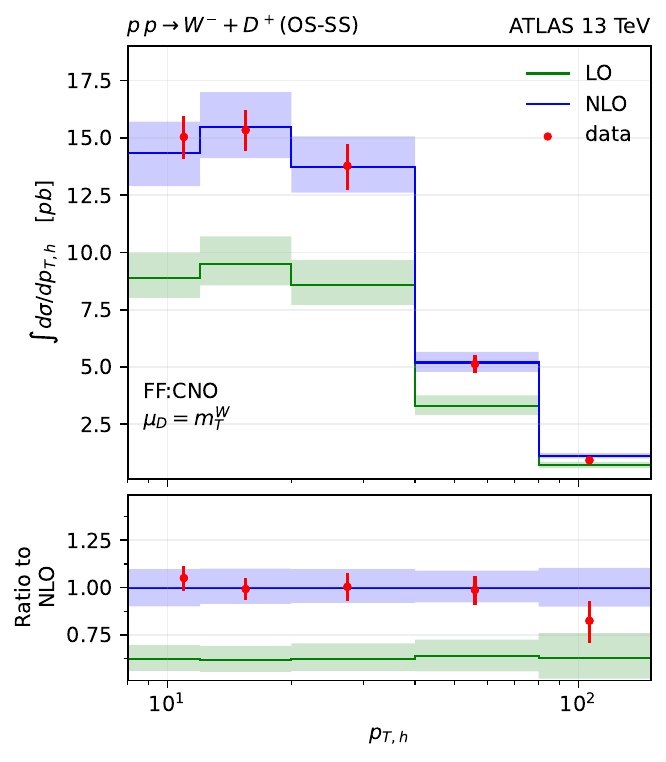}
	\includegraphics[width=65mm]{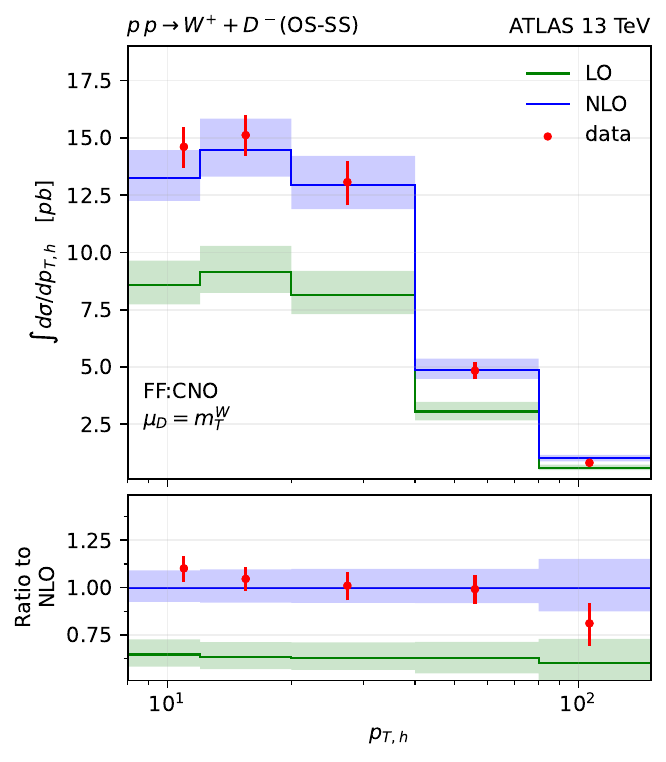}	
	\includegraphics[width=65mm]{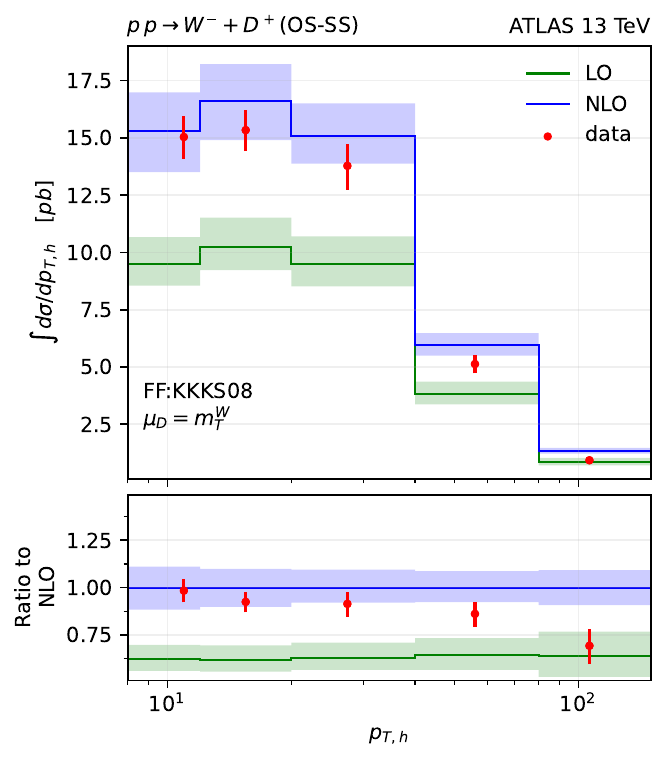}
	\includegraphics[width=65mm]{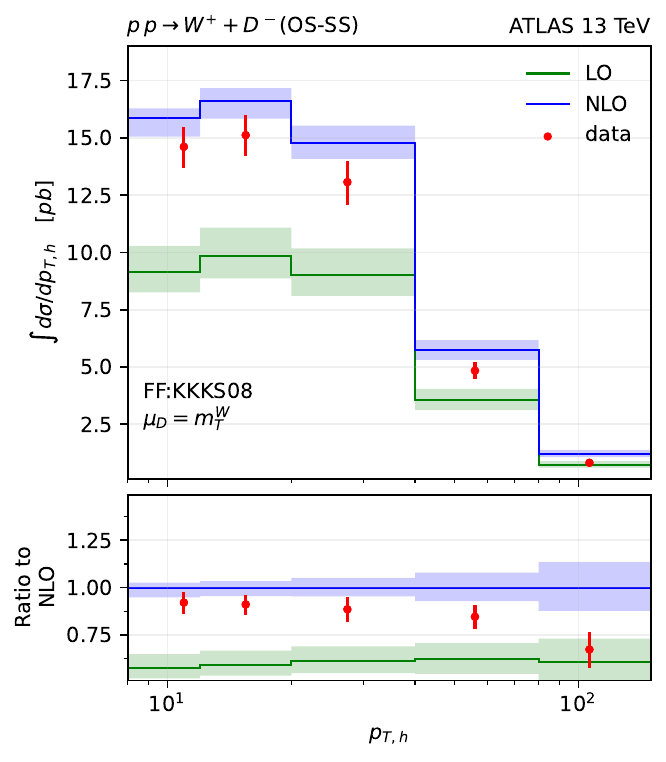}
	\caption{Comparison of LO (green) and NLO (blue) predictions with data (red). The $p_{T,h}$ distribution is integrated over each differential bin and is shown for all the $D$-hadron cases. The top plots have been produced using the CNO fragmentation function, while the bottom plots using  KKKS08. The left and right columns show the results for $W^-$ and $W^+$, respectively. All include the OS$-$SS prescription.}
	\label{fig:wd_pthad}
\end{figure}
\begin{figure}[tp]
	\centering  
	\includegraphics[width=65mm]{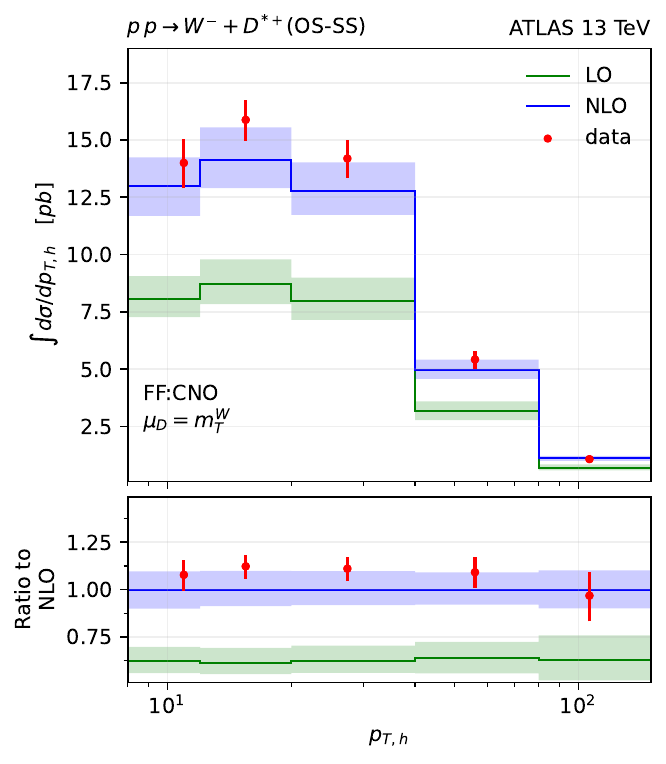}
	\includegraphics[width=65mm]{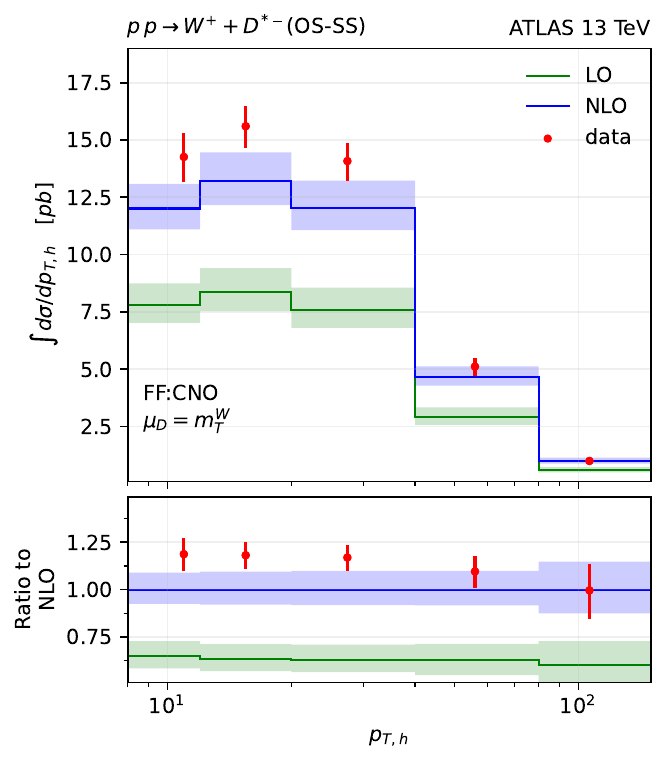}	
	\includegraphics[width=65mm]{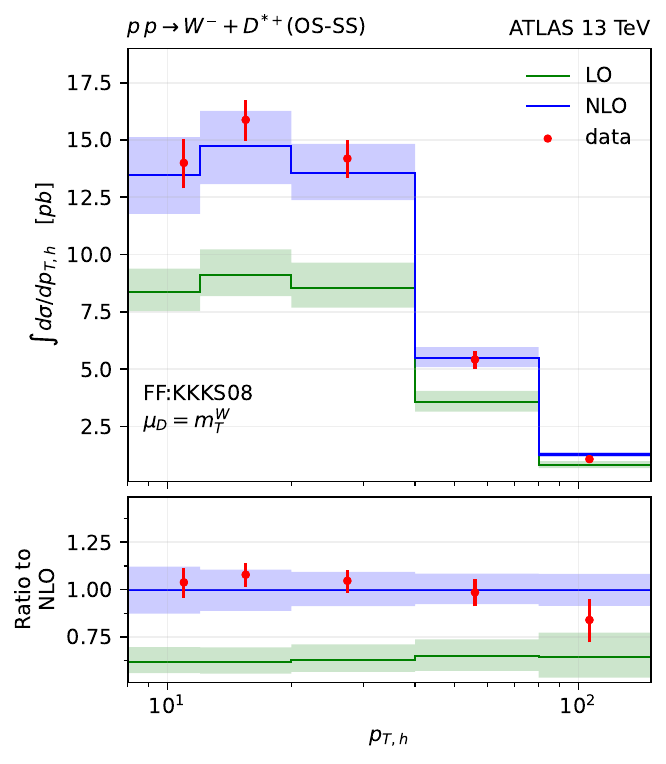}
	\includegraphics[width=65mm]{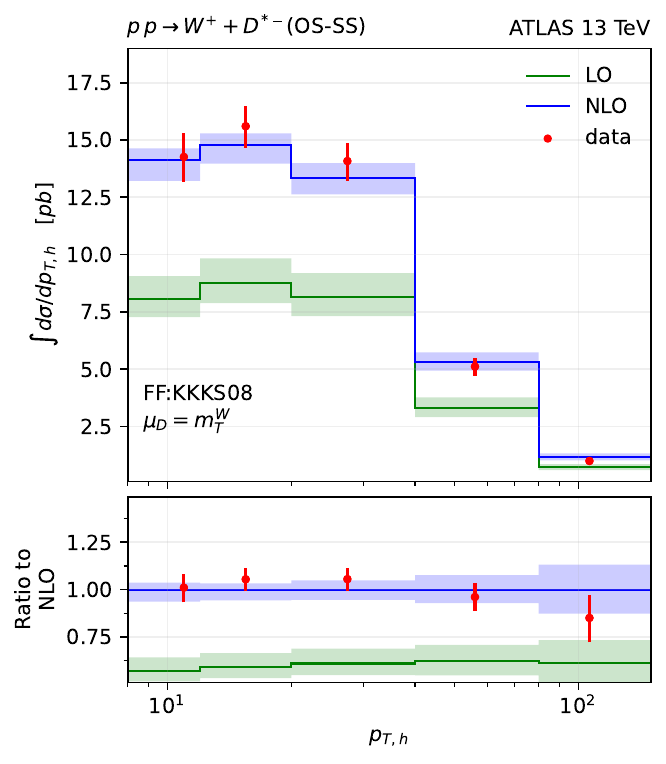}
	\caption{Comparison of LO (green) and NLO (blue) predictions with data (red). The $p_{T,h}$ distribution is integrated over each differential bin and is shown for all the $D^*$-hadron cases. The top plots have been produced using the CNO fragmentation function, while the bottom plots using  KKKS08. The left and right columns show the results for $W^-$ and $W^+$, respectively. All include the OS$-$SS prescription.}
	\label{fig:wdstar_pthad}
\end{figure}
In addition to the lepton rapidity $|\eta_\ell|$, we also consider predictions differential in $p_{T,h}$ both for $W$-boson plus $D$-hadron and $W$-boson plus $D^{*}$-hadron production. They are shown in Fig.~\ref{fig:wd_pthad} and Fig.~\ref{fig:wdstar_pthad}, respectively, with the same format as in Figs.~\ref{fig:wd_yl} and \ref{fig:wdstar_yl}.

Overall, we note that also the $p_{T,h}$ distributions are in good agreement with the data.
In the $D$-hadron case shown in Fig.~\ref{fig:wd_pthad}, the NLO correction is about a $40\%$ in the whole $p_{T,h}$ region. It is interesting to notice that the results obtained with the CNO fragmentation function are compatible with the data in the full spectrum, while we observe disagreement in the high-$p_{T,h}$ region for the predictions with KKKS08.
%
The $p_{T,h}$ distribution for the $D^*$-hadron case is shown in Fig.~\ref{fig:wdstar_pthad} and similarly to the $D$-hadron case features a NLO correction of $40\%$.
Here, predictions with both FF sets undershoot the data points in the low-$p_{T,h}$ region, however remaining generally compatible within uncertainties.
We again observe some disagreement in the high-$p_{T,h}$ region when the KKKS08 set is adopted.

\begin{figure}[tp]
	\centering  
	\includegraphics[scale=0.45]{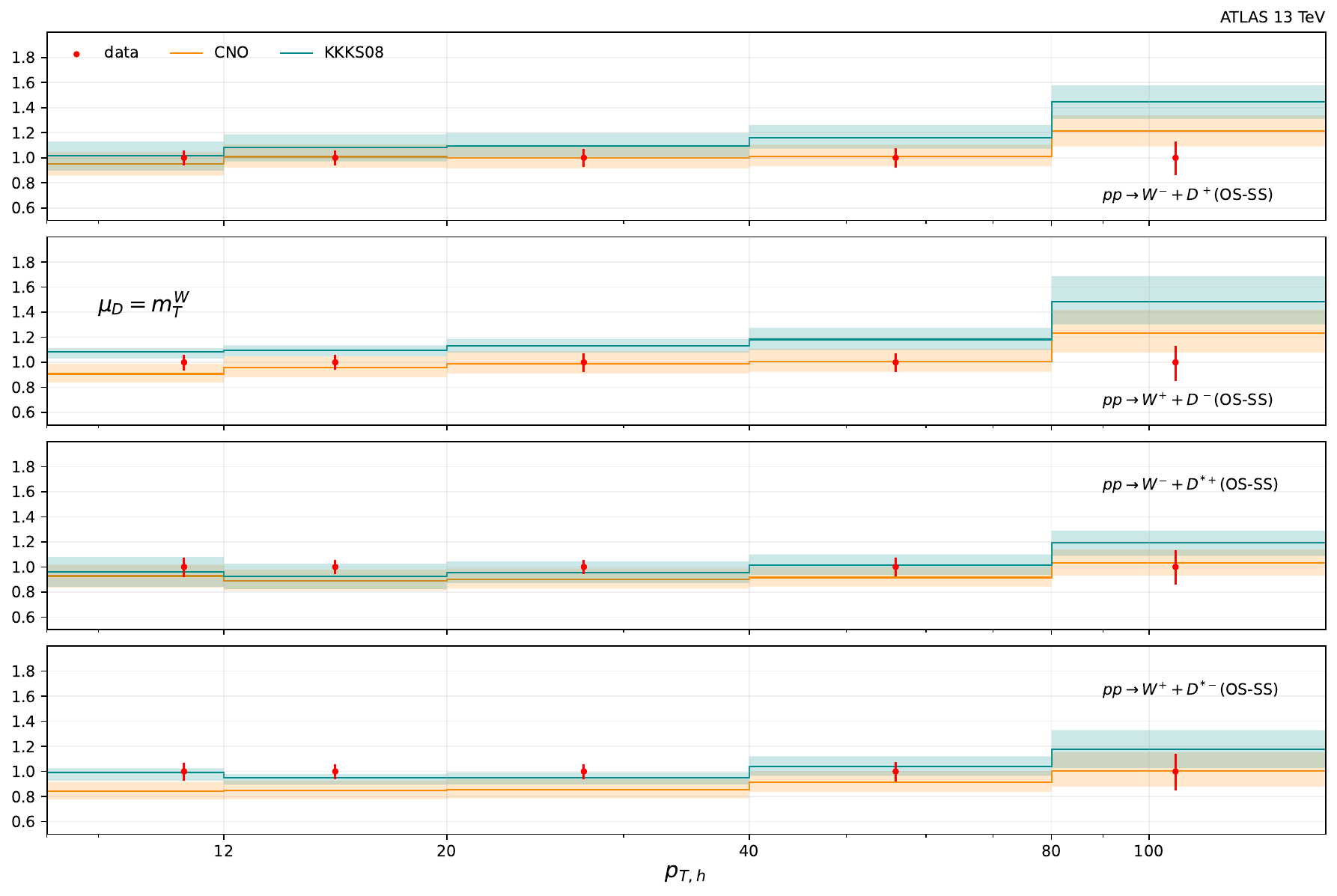}
	\caption{Comparison of the NLO results obtained for the different sets of fragmentation functions shown in Fig.~\ref{fig:wd_pthad} and Fig.~\ref{fig:wdstar_pthad} The plots corresponds to the ratio to data of the $p_{T, h}$ distribution for $D$ and $D^*$-hadrons.}
	\label{fig:squeezed_pthad}
\end{figure}
A direct comparison of the $p_{T,h}$ distribution obtained with the two sets of fragmentation functions is shown in Fig.~\ref{fig:squeezed_pthad}. From this plots it is easier to notice that the $p_{T,h}$ distribution is better described by the prediction obtained using the CNO fragmentation function in the low $p_{T,h}$ region, while the results obtained with KKKS08 are closer to data in the high-$p_{T,h}$ region, especially for the $D^{*}$-hadron cases. However, it is interesting to notice that the difference between the two seems to be just a shift that does not affect the shape of the distributions.

\section{Conclusions}
\label{sec:conclusions}

In this paper, we have provided theoretical predictions for observables related to identified hadron production at LHC.
In the first part, we have described how the antenna formalism has been extended to deal with infrared singularities associated to identified parton production in the short-distance cross section.
We have detailed the analytical ingredients necessary for the computation of observables related to the production of a hadron in association with an electroweak boson decaying leptonically at NLO accuracy.
In the second part, we have performed a detailed comparison 
between our predictions and experimental data for both the production of a $Z$ boson in association with charged hadrons inside a jet, as measured by the LHCb collaboration~\cite{LHCb:2022rky}, as well as the production of a $W$ boson in association with charmed hadrons, as measured by the ATLAS collaboration in~\cite{ATLAS:2023ibp}.
In our predictions, we have considered various choices of light- or heavy-quark fragmentation functions.

In observables related to the production of a hadron inside a jet in association with a $Z$-boson, we have found that our results depend significantly on the choice of the fragmentation function, highlighting the importance of including this data in FF fits.
The agreement of theory predictions with experimental data is also largely determined by the kinematical region considered, as for low values of the longitudinal momentum fraction of the hadron inside the jet we cannot expect perturbative fixed-order calculations to provide a fair description of data.
It would be interesting to match our NLO calculations with resummation, possibly including non-perturbative modelling, in order to enlarge the range of validity of our theoretical predictions.

Predictions for $W$-boson production in association with a charmed hadron have instead been found to be less dependent on the heavy quark fragmentation function adopted.
We have shown that two rather different FF sets (CNO and KKKS08) give similar sizeable NLO corrections, which bring theory predictions in good agreement with data, for both the lepton rapidity $|\eta_\ell|$ and the hadron transverse momentum $p_{T,h}$ distributions.
Both CNO and KKKS08 sets contain parameters that have been fitted to $\mathrm{e}^+\mathrm{e}^-$ data and it is reassuring to observe that these FF sets are also able to provide a reasonable description of LHC data.

Of course, it would be desirable to push our theory predictions to NNLO and the work presented in this paper is a first step towards this direction.
Consistent NNLO fixed-order predictions would also require the availability of NNLO heavy quark fragmentation functions. Within the perturbative fragmentation function framework,
choices related to the delicate treatment of non-perturbative effects in association with soft gluon resummation have been shown to give rather different results~\cite{Bonino:2023icn}. This behaviour is also enhanced by the small value of the charm mass, which is barely above the non-perturbative energy scale.
Hence, pushing the accuracy of predictions for identified hadron production at the LHC to NNLO will require not only technical work within the antenna subtraction formalism but also improvements on the conceptual side to better understand the heavy quark fragmentation dynamics.

\begin{acknowledgments}
We would like to thank Luca Rottoli for useful discussions on the hadron fragmentation functions 
belonging to the sets NNFF1.0 and NNFF1.1.  We are grateful to Leonardo Bonino for a careful reading of the manuscript.
This research was supported by the Swiss National Science Foundation (SNF) under contract 200021-197130.   
Numerical simulations were facilitated by the HighPerformance Computing group at ETH Z\"urich and the Swiss National
Supercomputing Centre (CSCS) under project ID ETH5f.
\end{acknowledgments}

%
\bibliography{V_hfrag}

\providecommand{\href}[2]{#2}\begingroup\raggedright\begin{thebibliography}{10}

\bibitem{Altarelli:1977zs}
G.~Altarelli and G.~Parisi, \emph{{Asymptotic Freedom in Parton Language}},
  \href{https://doi.org/10.1016/0550-3213(77)90384-4}{\emph{Nucl. Phys. B}
  {\bfseries 126} (1977) 298}.

\bibitem{Aversa:1988vb}
F.~Aversa, P.~Chiappetta, M.~Greco and J.P.~Guillet, \emph{{QCD Corrections to
  Parton-Parton Scattering Processes}},
  \href{https://doi.org/10.1016/0550-3213(89)90288-5}{\emph{Nucl. Phys. B}
  {\bfseries 327} (1989) 105}.

\bibitem{Rijken:1996ns}
P.J.~Rijken and W.L.~van Neerven, \emph{{Higher order QCD corrections to the
  transverse and longitudinal fragmentation functions in electron - positron
  annihilation}},
  \href{https://doi.org/10.1016/S0550-3213(96)00669-4}{\emph{Nucl. Phys. B}
  {\bfseries 487} (1997) 233}
  [\href{https://arxiv.org/abs/hep-ph/9609377}{{\ttfamily hep-ph/9609377}}].

\bibitem{Mitov:2006ic}
A.~Mitov, S.~Moch and A.~Vogt, \emph{{Next-to-Next-to-Leading Order Evolution
  of Non-Singlet Fragmentation Functions}},
  \href{https://doi.org/10.1016/j.physletb.2006.05.005}{\emph{Phys. Lett. B}
  {\bfseries 638} (2006) 61}
  [\href{https://arxiv.org/abs/hep-ph/0604053}{{\ttfamily hep-ph/0604053}}].

\bibitem{Bonino:2024qbh}
L.~Bonino, T.~Gehrmann and G.~Stagnitto, \emph{{Semi-inclusive deep-inelastic
  scattering at NNLO in QCD}},
  \href{https://arxiv.org/abs/2401.16281}{{\ttfamily 2401.16281}}.

\bibitem{ATLAS:2014jkm}
{\scshape ATLAS} collaboration, \emph{{Measurement of the production of a $W$
  boson in association with a charm quark in $pp$ collisions at $\sqrt{s} =$ 7
  TeV with the ATLAS detector}},
  \href{https://doi.org/10.1007/JHEP05(2014)068}{\emph{JHEP} {\bfseries 05}
  (2014) 068} [\href{https://arxiv.org/abs/1402.6263}{{\ttfamily 1402.6263}}].

\bibitem{CMS:2018dxg}
{\scshape CMS} collaboration, \emph{{Measurement of associated production of a
  W boson and a charm quark in proton-proton collisions at $\sqrt{s} =$ 13
  TeV}}, \href{https://doi.org/10.1140/epjc/s10052-019-6752-1}{\emph{Eur. Phys.
  J. C} {\bfseries 79} (2019) 269}
  [\href{https://arxiv.org/abs/1811.10021}{{\ttfamily 1811.10021}}].

\bibitem{CMS:2021oxn}
{\scshape CMS} collaboration, \emph{{Measurements of the associated production
  of a W boson and a charm quark in proton\textendash{}proton collisions at
  $\sqrt{s}=8$ TeV }},
  \href{https://doi.org/10.1140/epjc/s10052-022-10897-7}{\emph{Eur. Phys. J. C}
  {\bfseries 82} (2022) 1094}
  [\href{https://arxiv.org/abs/2112.00895}{{\ttfamily 2112.00895}}].

\bibitem{CMS:2023aim}
{\scshape CMS} collaboration, \emph{{Measurement of the production cross
  section for a W boson in association with a charm quark in
  proton\textendash{}proton collisions at $\sqrt{s} = 13\,\hbox {TeV}$}},
  \href{https://doi.org/10.1140/epjc/s10052-023-12258-4}{\emph{Eur. Phys. J. C}
  {\bfseries 84} (2024) 27} [\href{https://arxiv.org/abs/2308.02285}{{\ttfamily
  2308.02285}}].

\bibitem{ATLAS:2023ibp}
{\scshape ATLAS} collaboration, \emph{{Measurement of the production of a $W$
  boson in association with a charmed hadron in $pp$ collisions at $\sqrt{s} =
  13\,\mathrm{TeV}$ with the ATLAS detector}},
  \href{https://doi.org/10.1103/PhysRevD.108.032012}{\emph{Phys. Rev. D}
  {\bfseries 108} (2023) 032012}
  [\href{https://arxiv.org/abs/2302.00336}{{\ttfamily 2302.00336}}].

\bibitem{LHCb:2022rky}
{\scshape LHCb} collaboration, \emph{{Multidifferential study of identified
  charged hadron distributions in $Z$-tagged jets in proton-proton collisions
  at $\sqrt{s}=$13 TeV}},
  \href{https://doi.org/10.1103/PhysRevD.108.L031103}{\emph{Phys. Rev. D}
  {\bfseries 108} (2023) L031103}
  [\href{https://arxiv.org/abs/2208.11691}{{\ttfamily 2208.11691}}].

\bibitem{Catani:1996vz}
S.~Catani and M.H.~Seymour, \emph{{A General algorithm for calculating jet
  cross-sections in NLO QCD}},
  \href{https://doi.org/10.1016/S0550-3213(96)00589-5}{\emph{Nucl. Phys. B}
  {\bfseries 485} (1997) 291}
  [\href{https://arxiv.org/abs/hep-ph/9605323}{{\ttfamily hep-ph/9605323}}].

\bibitem{Frederix:2018nkq}
R.~Frederix, S.~Frixione, V.~Hirschi, D.~Pagani, H.S.~Shao and M.~Zaro,
  \emph{{The automation of next-to-leading order electroweak calculations}},
  \href{https://doi.org/10.1007/JHEP11(2021)085}{\emph{JHEP} {\bfseries 07}
  (2018) 185} [\href{https://arxiv.org/abs/1804.10017}{{\ttfamily
  1804.10017}}].

\bibitem{Czakon:2021ohs}
M.L.~Czakon, T.~Generet, A.~Mitov and R.~Poncelet, \emph{{B-hadron production
  in NNLO QCD: application to LHC t$ \overline{t} $ events with leptonic
  decays}}, \href{https://doi.org/10.1007/JHEP10(2021)216}{\emph{JHEP}
  {\bfseries 10} (2021) 216}
  [\href{https://arxiv.org/abs/2102.08267}{{\ttfamily 2102.08267}}].

\bibitem{Gehrmann:2022cih}
T.~Gehrmann and R.~Sch\"urmann, \emph{Photon fragmentation in the antenna
  subtraction formalism},
  \href{https://doi.org/10.1007/JHEP04(2022)031}{\emph{JHEP} {\bfseries 04}
  (2022) 031} [\href{https://arxiv.org/abs/2201.06982}{{\ttfamily
  2201.06982}}].

\bibitem{Chen:2022gpk}
X.~Chen, T.~Gehrmann, E.W.N.~Glover, M.~H\"ofer, A.~Huss and R.~Sch\"urmann,
  \emph{{Single photon production at hadron colliders at NNLO QCD with
  realistic photon isolation}},
  \href{https://doi.org/10.1007/JHEP08(2022)094}{\emph{JHEP} {\bfseries 08}
  (2022) 094} [\href{https://arxiv.org/abs/2205.01516}{{\ttfamily
  2205.01516}}].

\bibitem{Gehrmann:2022pzd}
T.~Gehrmann and G.~Stagnitto, \emph{{Antenna subtraction at NNLO with
  identified hadrons}},
  \href{https://doi.org/10.1007/JHEP10(2022)136}{\emph{JHEP} {\bfseries 10}
  (2022) 136} [\href{https://arxiv.org/abs/2208.02650}{{\ttfamily
  2208.02650}}].

\bibitem{Alwall:2014hca}
J.~Alwall, R.~Frederix, S.~Frixione, V.~Hirschi, F.~Maltoni, O.~Mattelaer
  et~al., \emph{{The automated computation of tree-level and next-to-leading
  order differential cross sections, and their matching to parton shower
  simulations}}, \href{https://doi.org/10.1007/JHEP07(2014)079}{\emph{JHEP}
  {\bfseries 07} (2014) 079} [\href{https://arxiv.org/abs/1405.0301}{{\ttfamily
  1405.0301}}].

\bibitem{Liu:2023fsq}
C.~Liu, X.~Shen, B.~Zhou and J.~Gao, \emph{{Automated calculation of jet
  fragmentation at NLO in QCD}},
  \href{https://doi.org/10.1007/JHEP09(2023)108}{\emph{JHEP} {\bfseries 09}
  (2023) 108} [\href{https://arxiv.org/abs/2305.14620}{{\ttfamily
  2305.14620}}].

\bibitem{Bevilacqua:2021ovq}
G.~Bevilacqua, M.V.~Garzelli, A.~Kardos and L.~Toth, \emph{{W + charm
  production with massive c quarks in PowHel}},
  \href{https://doi.org/10.1007/JHEP04(2022)056}{\emph{JHEP} {\bfseries 04}
  (2022) 056} [\href{https://arxiv.org/abs/2106.11261}{{\ttfamily
  2106.11261}}].

\bibitem{FerrarioRavasio:2023kjq}
S.~Ferrario~Ravasio and C.~Oleari, \emph{{NLO + parton-shower generator for Wc
  production in the POWHEG BOX RES}},
  \href{https://doi.org/10.1140/epjc/s10052-023-11853-9}{\emph{Eur. Phys. J. C}
  {\bfseries 83} (2023) 684}
  [\href{https://arxiv.org/abs/2304.13791}{{\ttfamily 2304.13791}}].

\bibitem{Currie:2013vh}
J.~Currie, E.W.N.~Glover and S.~Wells, \emph{{Infrared Structure at NNLO Using
  Antenna Subtraction}},
  \href{https://doi.org/10.1007/JHEP04(2013)066}{\emph{JHEP} {\bfseries 04}
  (2013) 066} [\href{https://arxiv.org/abs/1301.4693}{{\ttfamily 1301.4693}}].

\bibitem{Daleo:2006xa}
A.~Daleo, T.~Gehrmann and D.~Maitre, \emph{{Antenna subtraction with hadronic
  initial states}},
  \href{https://doi.org/10.1088/1126-6708/2007/04/016}{\emph{JHEP} {\bfseries
  04} (2007) 016} [\href{https://arxiv.org/abs/hep-ph/0612257}{{\ttfamily
  hep-ph/0612257}}].

\bibitem{Catani:1998bh}
S.~Catani, \emph{{The Singular behavior of QCD amplitudes at two loop order}},
  \href{https://doi.org/10.1016/S0370-2693(98)00332-3}{\emph{Phys. Lett. B}
  {\bfseries 427} (1998) 161}
  [\href{https://arxiv.org/abs/hep-ph/9802439}{{\ttfamily hep-ph/9802439}}].

\bibitem{LHCb:2021stx}
{\scshape LHCb} collaboration, \emph{{Study of Z Bosons Produced in Association
  with Charm in the Forward Region}},
  \href{https://doi.org/10.1103/PhysRevLett.128.082001}{\emph{Phys. Rev. Lett.}
  {\bfseries 128} (2022) 082001}
  [\href{https://arxiv.org/abs/2109.08084}{{\ttfamily 2109.08084}}].

\bibitem{Cacciari:2008gp}
M.~Cacciari, G.P.~Salam and G.~Soyez, \emph{{The anti-$k_t$ jet clustering
  algorithm}}, \href{https://doi.org/10.1088/1126-6708/2008/04/063}{\emph{JHEP}
  {\bfseries 04} (2008) 063} [\href{https://arxiv.org/abs/0802.1189}{{\ttfamily
  0802.1189}}].

\bibitem{NNPDF:2017mvq}
{\scshape NNPDF} collaboration, \emph{{Parton distributions from high-precision
  collider data}},
  \href{https://doi.org/10.1140/epjc/s10052-017-5199-5}{\emph{Eur. Phys. J. C}
  {\bfseries 77} (2017) 663}
  [\href{https://arxiv.org/abs/1706.00428}{{\ttfamily 1706.00428}}].

\bibitem{Buckley:2014ana}
A.~Buckley, J.~Ferrando, S.~Lloyd, K.~Nordstr\"om, B.~Page, M.~R\"ufenacht
  et~al., \emph{{LHAPDF6: parton density access in the LHC precision era}},
  \href{https://doi.org/10.1140/epjc/s10052-015-3318-8}{\emph{Eur. Phys. J. C}
  {\bfseries 75} (2015) 132} [\href{https://arxiv.org/abs/1412.7420}{{\ttfamily
  1412.7420}}].

\bibitem{Binnewies:1995pt}
J.~Binnewies, B.A.~Kniehl and G.~Kramer, \emph{{Pion and kaon production in e+
  e- and e p collisions at next-to-leading order}},
  \href{https://doi.org/10.1103/PhysRevD.52.4947}{\emph{Phys. Rev. D}
  {\bfseries 52} (1995) 4947}
  [\href{https://arxiv.org/abs/hep-ph/9503464}{{\ttfamily hep-ph/9503464}}].

\bibitem{Bertone:2018ecm}
{\scshape NNPDF} collaboration, \emph{{Charged hadron fragmentation functions
  from collider data}},
  \href{https://doi.org/10.1140/epjc/s10052-018-6130-4}{\emph{Eur. Phys. J. C}
  {\bfseries 78} (2018) 651}
  [\href{https://arxiv.org/abs/1807.03310}{{\ttfamily 1807.03310}}].

\bibitem{deFlorian:2007aj}
D.~de~Florian, R.~Sassot and M.~Stratmann, \emph{{Global analysis of
  fragmentation functions for pions and kaons and their uncertainties}},
  \href{https://doi.org/10.1103/PhysRevD.75.114010}{\emph{Phys. Rev. D}
  {\bfseries 75} (2007) 114010}
  [\href{https://arxiv.org/abs/hep-ph/0703242}{{\ttfamily hep-ph/0703242}}].

\bibitem{deFlorian:2007ekg}
D.~de~Florian, R.~Sassot and M.~Stratmann, \emph{{Global analysis of
  fragmentation functions for protons and charged hadrons}},
  \href{https://doi.org/10.1103/PhysRevD.76.074033}{\emph{Phys. Rev. D}
  {\bfseries 76} (2007) 074033}
  [\href{https://arxiv.org/abs/0707.1506}{{\ttfamily 0707.1506}}].

\bibitem{Bertone:2017tyb}
{\scshape NNPDF} collaboration, \emph{{A determination of the fragmentation
  functions of pions, kaons, and protons with faithful uncertainties}},
  \href{https://doi.org/10.1140/epjc/s10052-017-5088-y}{\emph{Eur. Phys. J. C}
  {\bfseries 77} (2017) 516}
  [\href{https://arxiv.org/abs/1706.07049}{{\ttfamily 1706.07049}}].

\bibitem{Gao:2024nkz}
J.~Gao, C.~Liu, X.~Shen, H.~Xing and Y.~Zhao, \emph{{Simultaneous Determination
  of Fragmentation Functions and Test on Momentum Sum Rule}},
  \href{https://arxiv.org/abs/2401.02781}{{\ttfamily 2401.02781}}.

\bibitem{ParticleDataGroup:2020ssz}
{\scshape Particle Data Group} collaboration, \emph{{Review of Particle
  Physics}}, \href{https://doi.org/10.1093/ptep/ptaa104}{\emph{PTEP} {\bfseries
  2020} (2020) 083C01}.

\bibitem{Cacciari:2005uk}
M.~Cacciari, P.~Nason and C.~Oleari, \emph{{A Study of heavy flavored meson
  fragmentation functions in e+ e- annihilation}},
  \href{https://doi.org/10.1088/1126-6708/2006/04/006}{\emph{JHEP} {\bfseries
  04} (2006) 006} [\href{https://arxiv.org/abs/hep-ph/0510032}{{\ttfamily
  hep-ph/0510032}}].

\bibitem{Kneesch:2007ey}
T.~Kneesch, B.A.~Kniehl, G.~Kramer and I.~Schienbein, \emph{{Charmed-meson
  fragmentation functions with finite-mass corrections}},
  \href{https://doi.org/10.1016/j.nuclphysb.2008.02.015}{\emph{Nucl. Phys. B}
  {\bfseries 799} (2008) 34} [\href{https://arxiv.org/abs/0712.0481}{{\ttfamily
  0712.0481}}].

\bibitem{Mele:1990yq}
B.~Mele and P.~Nason, \emph{{Next-to-leading QCD calculation of the heavy quark
  fragmentation function}},
  \href{https://doi.org/10.1016/0370-2693(90)90704-A}{\emph{Phys. Lett. B}
  {\bfseries 245} (1990) 635}.

\bibitem{Gehrmann-DeRidder:2023gdl}
A.~Gehrmann-De~Ridder, T.~Gehrmann, E.W.N.~Glover, A.~Huss, A.R.~Garcia and
  G.~Stagnitto, \emph{{Precise QCD predictions for W-boson production in
  association with a charm jet}},
  \href{https://doi.org/10.1140/epjc/s10052-024-12715-8}{\emph{Eur. Phys. J. C}
  {\bfseries 84} (2024) 361}
  [\href{https://arxiv.org/abs/2311.14991}{{\ttfamily 2311.14991}}].

\bibitem{Bonino:2023icn}
L.~Bonino, M.~Cacciari and G.~Stagnitto, \emph{{Heavy Quark Fragmentation in
  $e^+e^-$ Collisions to NNLO+NNLL Accuracy in Perturbative QCD}},
  \href{https://arxiv.org/abs/2312.12519}{{\ttfamily 2312.12519}}.

\end{thebibliography}\endgroup

\end{document}